\pdfoutput=1
\PassOptionsToPackage{
pdfencoding=auto,
pdfnewwindow=true,
pdfusetitle=false,
bookmarks=true,
bookmarksnumbered=true,
bookmarksopen=true,
pdfpagemode=UseThumbs,
bookmarksopenlevel=1,
pdfpagelabels=false,
breaklinks=true,  
backref=false, 
colorlinks=true,
}{hyperref}
\RequirePackage{etex}
\PassOptionsToPackage{skip=0pt, font=small,labelfont=bf}{caption}
\documentclass[10pt,showpacs,letterpaper,aps,pra,twocolumn,nofootinbib,superscriptaddress,floatfix]{revtex4-2} 
\usepackage{hyperref}

\usepackage{amsthm,amsmath,amssymb,graphicx,comment,dsfont} 
\usepackage{color}
\usepackage{bbold}
\usepackage[none]{hyphenat}
\usepackage{fancyhdr}
\newcommand{\ket}[1]{\left| #1 \right\rangle}

\newcommand{\beq}{\begin{equation}}
\newcommand{\eeq}{\end{equation}}
\newcommand{\bea}{\begin{align}}
\newcommand{\eea}{\end{align}}

\usepackage{graphicx}
\usepackage[usenames,dvipsnames,table]{xcolor} 
\usepackage{mathtools}
\usepackage[normalem]{ulem}
\input{preamble}

\definecolor{googleblue}{RGB}{34, 0, 204}
\definecolor{panblue}{RGB}{0,24,150}
\definecolor{carmine}{RGB}{150, 0, 24}
\hypersetup{
unicode=true,
bookmarksnumbered=false,
bookmarksopen=false,
breaklinks=false,
colorlinks=true,
linkcolor=carmine,
citecolor=googleblue,
urlcolor=panblue,
anchorcolor=OliveGreen}

\allowdisplaybreaks

\usepackage{subcaption}
\usepackage[labelformat=simple]{subcaption}

\usepackage{microtype}
\microtypecontext{spacing=nonfrench}
\microtypesetup{
expansion={true,nocompatibility},
protrusion={true,nocompatibility},
activate={true,nocompatibility},
tracking=true,
kerning=true,
spacing={true}
}

\usepackage[all=normal,floats=tight,mathspacing=tight,wordspacing=tight,paragraphs=normal,tracking=tight,charwidths=tight,mathdisplays=normal,sections=normal,margins=normal]{savetrees}

\everypar=\expandafter{\the\everypar\loosness=-1 }
\newcommand{\nocontentsline}[3]{}
\let\oldaddcontentsline\addcontentsline
\newcommand{\tocless}[2]{%
  \let\addcontentsline=\nocontentsline#1{#2}
  \let\addcontentsline\oldaddcontentsline}

\begin{document}
\title{Addressing some common objections to generalized noncontextuality}
\author{David Schmid}
\affiliation{International Centre for Theory of Quantum Technologies, University of Gda\'nsk, 80-308 Gda\'nsk, Poland}
\email{davidschmid10@gmail.com}
\author{John H. Selby}
\email{john.h.selby@gmail.com}
\affiliation{International Centre for Theory of Quantum Technologies, University of Gda\'nsk, 80-308 Gda\'nsk, Poland}
\author{Robert W. Spekkens}
\email{rspekkens@perimeterinstitute.ca}
\affiliation{Perimeter Institute for Theoretical Physics, 31 Caroline Street North, Waterloo, Ontario Canada N2L 2Y5}

\date{\today}

\begin{abstract}
When should a given operational phenomenology be deemed to admit of a classical explanation? When it can be realized in a generalized-noncontextual ontological model. The case for answering the question in this fashion has been made in many previous works, and motivates research on the notion of generalized noncontextuality. Many criticisms and concerns have been raised, however, regarding the definition of this notion
 and of the possibility of testing it experimentally. In this work, we respond to some of the most common of these objections. One such objection is that the existence of a classical record of which laboratory procedure was actually performed in each run of an experiment implies that the operational equivalence relations that are a necessary ingredient of any proof of the failure of noncontextuality do not hold, and consequently that conclusions of nonclassicality based on these equivalences are mistaken. We explain why this concern in unfounded. Our response affords the opportunity for us to clarify certain facts about generalized noncontextuality, such as the possibility of having proofs of its failure based on a consideration of the subsystem structure of composite systems. Similarly, through our responses to each of the other objections, we elucidate some under-appreciated facts about the notion of generalized noncontextuality and experimental tests thereof.  
\end{abstract}

\maketitle
\tableofcontents

\section{Introduction}

The notion of generalized noncontextuality was introduced in Ref.~\cite{gencontext} as an extension of Kochen-Specker noncontextuality~\cite{KS}. 
Realizability by a generalized-noncontextual ontological model provides a notion of classical explainability for operational phenomena. Consequently, demonstrating that a given experiment {\em cannot} be explained within any generalized-noncontextual ontological model constitutes a rigorous proof of nonclassicality.
Many previous works have provided arguments for why generalized noncontextuality is a gold standard notion of classical explainability; see for instance Ref.~\cite{Leibniz}, or the introductions of Refs.~\cite{schmid2021guiding,selby2024linear}.
 We will touch on some of these arguments in passing in this work.

Our aim here, however, is to collect and respond to a number of {\em objections} that have been raised against the notion of generalized noncontextuality, including challenges to its motivations, its consistency, and its experimental testability.
We also elaborate on a number of other conceptual points that have the potential to be misunderstood, or points that are known to some experts but which we feel deserve wider recognition. 

Arguably, the most interesting analysis provided in this paper is the one in Sec.~\ref{sec:LabNotebook}, where we address 
the claim that proofs of contextuality are undermined by the existence of classical records of which operational procedures were carried out in each run of an experiment.
 We show that, contrary to this claim, 
 one can correctly assess the noncontextual-realizability of the operational statistics whether or not such records exist. Along the way, we demonstrate new possibilities for proving the failure of noncontextuality in scenarios with composite systems. 

Some topics that are {\em not} part of the scope of this article include:
(i) to provide an introduction to noncontextuality and the methods for testing or characterizing it, (ii) 
to provide an account of the arguments in favour of defining classical explainability of operational statistics in terms of realizability by a generalized-noncontextual ontological model, 
 and (iii) to discuss arguments concerning the relative merit  of generalized noncontextuality and Kochen-Specker noncontextuality.  We refer the reader to earlier works for these topics.

This paper assumes basic familiarity with the notions of operational theories, ontological models, and generalized noncontextuality. Where possible, we will focus on the simpler case of prepare-measure scenarios, although most of what is said can be generalized to scenarios with more general compositional 
structure~\cite{schmid2020structure}. 
Henceforth, the term ``noncontextual'' will be taken to refer to the notion of  generalized noncontextuality introduced in Ref.~\cite{gencontext}.

\section{Preliminaries}

 It is useful to distinguish two perspectives on witnessing the failure of generalized noncontextuality, which we shall refer to as {\em algebraic} and {\em geometric}. 
They provide two different ways of conceptualizing the constraints on the ontological model implied by operational identities under the assumption of noncontextuality.  In the algebraic approach, one seeks to determine whether one can represent each operational state by a probability distribution on the ontic state space, and each operational effect by a response function on the ontic state space, while respecting the identities that hold among these.   In the geometric approach, by contrast, one conceptualizes the identities holding among the operational states as stipulating the geometric shape of the convex hull of the states and similarly for the operational effects, and the question of ontological representability is expressed as a particular embedding for these geometric shapes.

The distinction should be understood in roughly the same way as the distinction between algebraic and geometric proofs of  the Kochen-Specker theorem\footnote{Algebraic proofs of the Kochen-Specker theorem proceed by considering a set of Hermitian operators (observables) and demonstrating that the functional relations that these satisfy cannot be satisfied by a set of classical variables when the value assigned to the variable representing a given observable is independent of what other observables are measured together with it.  Geometric proofs of the Kochen-Specker theorem, on the other hand, consider the orthogonality relations holding among rays in Hilbert space describing outcomes of a set of rank-1 projective measurements, and demonstrate that these rays cannot be assigned values 0 or 1 in such a way that a single element of every orthogonal set is assigned value 1, when the value assigned to a ray must be assigned independently of which orthogonal basis it is considered a part of.}, although in the case of Kochen-Specker, the arena for the geometric conditions is Hilbert space whereas for generalized noncontextuality it is the vector space of Hermitian operators (or, more generally, the vector space of GPT states and effects).  Just as any algebraic proof of the Kochen-Specker theorem can be translated into a geometric proof and vice-versa, so too can any algebraic approach to witnessing the failure of generalized noncontextuality be translated into the geometric approach and vice-versa. 

Although the difference between the approaches is a cosmetic one, 
sometimes one perspective or the other is more insightful or simple, so we recap both approaches in the next section. Of particular relevance to this work is the fact that, as we will see, the geometric perspective (which is the newer of the two) will often be useful for making {\em especially} clear how some past concerns about generalized noncontextuality are unfounded. Indeed, it seems likely to us that if this perspective had been adopted first, then many of these objections would never have arisen in the first place.

We will here focus on tests of generalized noncontextuality in prepare-measure scenarios. The generalization to arbitrary compositional scenarios can be found in Ref.~\cite{schmid2020structure}.

For a comprehensive introduction to noncontextuality (according to both perspectives we will discuss), we refer the reader to the series of three lectures at ~\cite{NCvid1,NCvid2,NCvid3} (and the references therein).

\subsection{Algebraic approach to witnessing the failure of generalized noncontextuality}\label{opequiv}

The algebraic approach was the first to be adopted for witnessing the failure of generalized noncontextuality~\cite{gencontext}  
and so is the more widely known of the two. In this approach, the relevant input data to the analysis are operational identities
---typically, linear constraints among the states and among the measurements. One uses these to derive noncontextuality inequalities, whose violation demonstrates that the operational predictions of the scenario cannot be reproduced by a noncontextual ontological model.

In the simplest experiment of interest, one implements a set of preparation procedures and a set of measurement procedures and one records the outcome statistics observed for each pairing. 
 An {\em operational state} is an operational equivalence class of preparation procedures, where two preparation procedures are defined to be operationally equivalent  if they generate the same statistics for all possible measurements.
An {\em operational effect} is an operational equivalence class of measurement-outcome pairs, where two such pairs are operationally equivalent if they are assigned the same probability by 
 all preparation procedures. 
The operational states generally satisfy nontrivial 
 identities, termed {\em operational identities}, as do the operational effects.  
A common form of such an identity for operational states is a linear dependence relation: 
\begin{equation} \label{linconstr}
\sum_{x\in X} \alpha_x \mathbf{s}_x = 0,
\end{equation}
where $\alpha_x \in \mathds{R}$ and $\mathbf{s}_x$ is an  operational state, represented as a vector in a GPT~\cite{Hardy,GPT_Barrett,chiribella2010probabilistic}.  In the case of quantum theory, these
    are simply representations of density operators in the real vector space of Hermitian operators, such as the Bloch vectors representing the density operators of a qubit. We will henceforth make frequent use of the GPT representation, and so we will often refer to operational states as {\em GPT states} and operational effects as {\em GPT effects}.

 An example of a circumstance implying a relation of the form of Eq.~\eqref{linconstr}
is when 
   a convex mixture of two GPT states is equal to a third GPT state.  Operational identities also hold among the GPT effects.  These identities can often be inferred by how a given state or effect is implemented (e.g., as a convex mixture of two others).  They can also be inferred from a tomographic characterization of the GPT states and GPT effects.  Finally, they can additionally be inferred from principles, such as no-signalling, or the absence of retrocausation.  
Demanding that these identities are also respected by the ontological representations of the states and effects implies 
constraints on the outcome statistics, typically in the form of inequalities known as noncontextuality inequalities.

\subsection{Geometric approach to witnessing the failure of generalized noncontextuality}
\label{gptnc}

The second perspective on noncontextuality is relatively recent. In this approach, the relevant input data to the analysis are a set of states and a set of measurement effects, as represented in some generalized probabilistic theory~\cite{Hardy,GPT_Barrett,chiribella2010probabilistic}. One then tests whether these can be embedded in a simplex and its dual (such that the probabilistic predictions are preserved); if such an embedding does not exist, this demonstrates that the operational statistics for that scenario cannot be reproduced within a noncontextual ontological model. 

More precisely: this approach relies on the fact that operational theories that are noncontextual are associated with generalized probabilistic theories that are simplex-embeddable~\cite{SchmidGPT,schmid2020structure}.  A generalized probabilistic theory, or GPT, is simplex-embeddable if its state space linearly embeds in a simplex and its effect space linearly embeds in the dual to that simplex, in such a way that the probabilities it predicts are unchanged. 
One can apply this approach to the study of particular scenarios and experiments as well. 
One simply obtains a characterization of the GPT states and effects realized in the experiment, termed an {\em accessible GPT fragment}~\cite{selby2023accessible}).  These characterizations provide inner bounds on the full GPT state space and the full GPT effect space respectively, such that if the accessible GPT fragment realized in the experiment is 
 not simplex-embeddable, then one can conclude that the GPT describing the system is not simplex-embeddable either. 

Thus, in this approach, one determines if a theory or experiment 
 is consistent with the principle of noncontextuality by testing whether the GPT representation of that theory or experiment
  satisfies a geometric criterion (simplex-embeddability). In this way, one need not consider operational identities as algebraic equations that in turn imply specific noncontextuality inequalities which one tests. 
 Rather,  one can think of the operational identities as constraints on the geometry of the state space.  For instance, the full set of operational identities holding among a set of states is simply a description of the geometry of the convex hull of those states. 

As noted above, although one {\em could} try to translate an analysis in one perspective to the other, 
one or the other approach will sometimes be more insightful.

\subsection{Operational identities involving subsystems} \label{opequivsub}

 Although the most commonly studied operational identities are of the form of Eq.~\eqref{linconstr}, there are other types  that can be leveraged for proving the failure of noncontextuallity.
 This was clearly stated even in the first paper on generalized noncontextuality, which noted (as just one other example) that distinct ways of purifying a given quantum state correspond to distinct but operationally equivalent preparation procedures in that state's operational equivalence class~\cite{gencontext}.
 For example, suppose that two GPT states on system $A$ are defined as
 \begin{align}
\mathbf{s}^{(1)}_A = \mathsf{tr}_B[\mathbf{s}^{(1)}_{AB}]\nonumber \\
\mathbf{s}^{(2)}_A = \mathsf{tr}_B[\mathbf{s}^{(2)}_{AB}],
\end{align}
 where $\mathsf{tr}_B$ is shorthand for the transformation $\mathbf{s}_{AB} \mapsto \mathbf{u}_
 B \cdot \mathbf{s}_{AB}$, 
 with $\mathbf{u}_B$ the unit effect for the system $B$.
 If it is the case that 
 \begin{equation}
\mathbf{s}^{(1)}_A = \mathbf{s}^{(2)}_A,
\end{equation}
then this describes a valid operational identity,
around which one could construct proofs of noncontextuality. 
One can also consider more general operational identities 
that involve both linear combinations and partial traces, e.g.,
\begin{equation}\label{genopeq}
\sum_{x\in X} \alpha_x \mathsf{tr}_B[\mathbf{s}^{(x)}_{AB}]=  \sum_{y\in Y} \beta_y \mathsf{tr}_B[\mathbf{s}^{(y)}_{AB}]. 
\end{equation}
We will not attempt to give a completely general algebraic description of the scope of operational identities one can consider; however, one can find a completely general diagrammatic description in Ref.~\cite{schmid2020structure,schmid2020unscrambling}. 

Most prior derivations of noncontextuality inequalities have relied on operational identities that are given by linear combinations like those in Eq.~\eqref{linconstr}---in particular, they did not make use of subsystem structure.  The only instance of a more general operational identity that we are aware of is in Ref.~\cite{determinism} (see Eq.~(8) and the surrounding discussion therein). 
In Section~\ref{sec:LabNotebook}, we will give a second example---a proof of contextuality that uses an operational identity of the form of Eq.~\eqref{genopeq}. While our example is quite simple, it demonstrates that one is generally {\em forced} to consider operational identities of this more general form if one wishes to determine the full implications of noncontextuality.

\subsection{Theory-agnostic tomography}\label{gpttomography}

A useful tool for determining the characterization of one's experiment within a generalized probabilistic theory is theory-agnostic tomography, also known as 
 GPT  tomography~\cite{mazurek2017experimentally,grabowecky2021experimentally}. While this tool is not required for understanding the definition of noncontextuality, theory-agnostic tomography is in many respects the ideal way of experimentally testing noncontextuality, and so it will be relevant to a number of the points we make herein. In theory-agnostic tomography, one carries out a large number of preparations and measurements on the given system, where these are chosen either randomly, or to roughly fill out an approximation of what one expects the true state and effect spaces to be. One does not assume anything a priori about the identity of each individual procedure (such as its GPT description), or about what GPT governs the experiment. Rather, one {\em extracts} (by an appropriate analysis) the GPT dimension and GPT descriptions of each state and effect in the experiment from the observed data. These realized GPT vectors then constitute inner approximations of the true GPT state and effect spaces. One can then use this information, for example, to assess whether the experiment is simplex-embeddable (classically-explainable) or not. 

\subsection{A standard proof of the failure of noncontextuality} \label{standardproof}

\begin{figure}[htbp!]
\begin{center}
\beq 
\begin{tikzpicture}
	\begin{pgfonlayer}{nodelayer}
		\node [style=none] (0) at (-0.5, -1) {};
		\node [style=none] (1) at (-0.5, 1) {};
		\node [style=none] (2) at (-1, -1) {};
		\node [style=none] (3) at (-1, -2) {};
		\node [style=none] (4) at (0, -2) {};
		\node [style=none] (5) at (0, -1) {};
		\node [style=none] (7) at (-0.5, 2) {};
		\node [style=none] (8) at (-1, 1) {};
		\node [style=none] (9) at (1, 1) {};
		\node [style=none] (10) at (1, 2) {};
		\node [style=none] (13) at (0.5, 2) {};
		\node [style=none] (14) at (0.5, 2.5) {};
		\node [style=none] (15) at (0.5, 0.5) {};
		\node [style=none] (16) at (0.5, 1) {};
		\node [style=none] (19) at (-0.5, -2.5) {};
		\node [style=none] (20) at (-0.5, -2) {};
		\node [style=right label] (21) at (-0.5, -2.25) {$X$};
		\node [style=right label] (23) at (0.5, 2.25) {$B$};
		\node [style=right label] (24) at (0.5, 0.75) {$Y$};
		\node [style=right label] (25) at (-0.5, 0) {$S$};
		\node [style=none] (26) at (-0.5, -1.5) {$P$};
		\node [style=none] (27) at (0, 1.5) {$M$};
		\node [style=none] (28) at (-0.5, -1) {};
		\node [style=none] (29) at (-0.5, -0.5) {};
	\end{pgfonlayer}
	\begin{pgfonlayer}{edgelayer}
		\draw [qWire] (1.center) to (0.center);
		\draw (2.center) to (5.center);
		\draw (5.center) to (4.center);
		\draw (4.center) to (3.center);
		\draw (3.center) to (2.center);
		\draw (7.center) to (10.center);
		\draw (10.center) to (9.center);
		\draw (9.center) to (8.center);
		\draw (8.center) to (7.center);
		\draw [cWire] (14.center) to (13.center);
		\draw [cWire] (16.center) to (15.center);
		\draw [cWire] (20.center) to (19.center);
	\end{pgfonlayer}
\end{tikzpicture}},
\eeq
\caption{The original PM scenario.}
\label{PMScenario}
\end{center}
\end{figure}
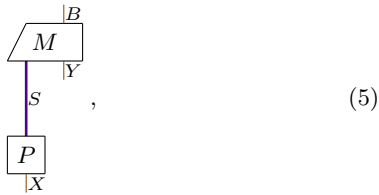

Consider a prepare-measure scenario, depicted in Figure~\ref{PMScenario}, defined by  a set  of GPT states indexed by the set $X$, $\{ {\bf s}_x\}_{x\in X}$, and a set of GPT measurements, indexed by the set $Y$ and where for each $y\in Y$, the GPT measurement is described by the set of effects $\{ {\bf e}_{b|y}\}_{b\in B}$. 
Imagine that the states satisfy some operational identities indexed by $j$, 
\begin{equation}\label{opidonS}
 \forall j : \sum_{x\in X} \alpha^{(j)}_x {\bf s}_x = 0
\end{equation}
for  $\alpha^{(j)}_x \in \mathds{R}$, so that the assumption of generalized noncontextuality implies  linear constraints 
of the same form on the associated epistemic states:
\begin{equation} \label{origopeq}
 \forall j:  \sum_{x\in X}  \alpha^{(j)}_x  \mu_x(\lambda) = 0 \quad \forall \ \lambda\in \Lambda,
\end{equation}
where $\mu_x$ is the epistemic state associated to the GPT state ${\bf s}_x$.
Similarly, we can imagine that the effects satisfy some operational identities indexed by $k$,
\beq
 \forall k: \sum_{b\in B, y\in Y}  \beta^{(k)}_{b,y} {\bf e}_{b|y} = 0,
\eeq 
for  $\beta^{(k)}_{b,y} \in \mathds{R}$, so that the assumption of generalized noncontextuality implies a linear constraint of the same form on the associated response functions:
\beq
 \forall k:  \sum_{b\in B, y\in Y}  \beta^{(k)}_{b,y}  \xi_{b|y}(\lambda) = 0 \quad \forall \ \lambda\in \Lambda,
\eeq 
where $\xi_{b|y}$ is the response function associated to the GPT effect ${\bf e}_{b|y}$.
Imagine moreover that one has derived  noncontextuality inequalities from these operational identities,
and that these have  been violated by the observed statistics in the experiment. In this case, one has found a proof of the failure of noncontextuality  in that prepare-measure scenario. 

\section{The lab notebook objection} \label{sec:LabNotebook}

A challenge that is sometimes made to the analysis given in the previous section is the following.  The choice of preparation in the experiment is typically recorded in the experimenter's lab notebook.  (Indeed, such a recording is {\em necessary} if the experimenter hopes to compute the statistics on which noncontextuality inequalities are tested.)
In particular, if the experiment includes two preparation procedures that are distinct but operationally equivalent, then which of these is implemented in a given run of the experiment is indicated in the lab notebook. Consequently, there {\em does} exist a measurement that distinguishes the two, namely, the measurement that reveals the physical state of the lab notebook.\footnote{Even if the experimenter does not take care to record which procedure was implemented, the environmental degrees of freedom within the laboratory are likely to carry away information about which it was (e.g., in the precise pattern of light rays scattered off the laboratory apparatus), and therefore these are likely to encode a record of which it was.}  According to this argument, therefore, no two preparation procedures are ever found to be operationally equivalent.  Because the assumption of generalized noncontextuality is an engine that turns operational equivalence relations among procedures into constraints on how they are represented in the ontological model, if there are no such equivalence relations, one obtains no constraints.  Hence, there is no opportunity to derive noncontextuality inequalities and thus no opportunity to discover a failure of noncontextuality.
 We will refer to this challenge as the ``lab notebook objection'' to generalized noncontextuality.

The first key fact that this objection misses is this: operational theories incorporate a notion of a physical system, which is treated as a primitive notion on which an individuating principle can be based. Preparation and measurement procedures are specific to a system, and consequently operational identities are evaluated  {\em relative to a system}.  Thus, for instance, two preparation procedures on a system $S$ are deemed to be operationally equivalent if they yield the same statistics for all measurements {\em on $S$}.   (For more general compositional scenarios, causal structure provides the individuating principle for procedures---see the discussion in Sec.~\ref{NCwrtcausalstr}.)

The system that is being prepared and measured in a given experiment can be conceptualized as the thing that acts as a causal mediary between the preparation device and the measurement device.  In an experiment wherein the preparations and measurements relate to the polarization of a photon, for instance, this degree of freedom constitutes $S$, the causal mediary.   The lab notebook in such an experiment is explicitly presumed {\em not} to act as such a causal mediary.  To imagine that the causal influence from the choice of preparation procedure to the outcome of the measurement is not mediated by the polarization degree of freedom of the photon, but rather by some physical records of how the preparation device was implemented is a radical and a priori rather implausible hypothesis about the causal structure of the experiment.  To put it more strongly: as long as one grants that the lab notebook is an independent physical system from $S$, assessments of  nonclassicality for system $S$ alone (as opposed to assessments for the joint system comprised of {\em the lab notebook together with $S$}) are based on  operational equivalences that are {\em defined} relative to measurements on $S$ alone.

Still, a stubborn skeptic might remain concerned about the case where one rather chooses to study the nonclassicality of the joint system defined by the lab notebook together with system $S$. Indeed, it has sometimes been claimed that in this case, one reaches a different verdict regarding the nonclassicality of the system---one that is inconsistent with the verdict obtained when the system $S$ alone is taken to be the system of interest.

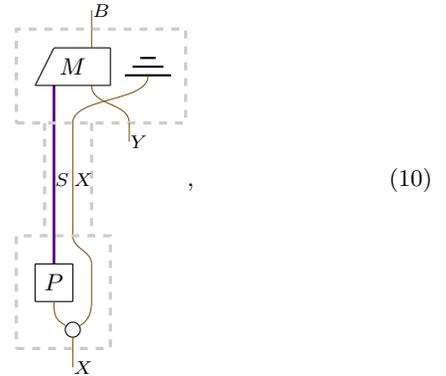
\begin{figure}[htbp!]
\begin{center}
\beq 
\begin{tikzpicture}
	\begin{pgfonlayer}{nodelayer}
		\node [style=none] (0) at (-0.5, -2.25) {};
		\node [style=none] (1) at (-0.5, 2.5) {};
		\node [style=none] (2) at (-1, -2.25) {};
		\node [style=none] (3) at (-1, -3.25) {};
		\node [style=none] (4) at (0, -3.25) {};
		\node [style=none] (5) at (0, -2.25) {};
		\node [style=none] (7) at (-0.5, 3.5) {};
		\node [style=none] (8) at (-1, 2.5) {};
		\node [style=none] (9) at (1, 2.5) {};
		\node [style=none] (10) at (1, 3.5) {};
		\node [style=none] (13) at (0.5, 3.5) {};
		\node [style=none] (14) at (0.5, 4.5) {};
		\node [style=none] (15) at (1.5, 1.5) {};
		\node [style=none] (16) at (0.5, 2.5) {};
		\node [style=white dot] (19) at (0, -4) {};
		\node [style=none] (20) at (-0.5, -3.25) {};
		\node [style=right label] (21) at (0, -5) {$X$};
		\node [style=right label] (23) at (0.5, 4.5) {$B$};
		\node [style=right label] (24) at (1.5, 1) {$Y$};
		\node [style=right label] (25) at (-0.5, 0) {$S$};
		\node [style=none] (26) at (-0.5, -2.75) {$P$};
		\node [style=none] (27) at (0, 3) {$M$};
		\node [style=none] (33) at (0, -5) {};
		\node [style=none] (34) at (0.5, -3.25) {};
		\node [style=none] (35) at (0.5, -2.25) {};
		\node [style=none] (36) at (0, -1.5) {};
		\node [style=right label] (38) at (0, 0) {$X$};
		\node [style=none] (40) at (0, 1.5) {};
		\node [style=none] (42) at (2, 2.75) {};
		\node [style=upground] (44) at (2, 3) {};
		\node [style=none] (45) at (-1.5, 4) {};
		\node [style=none] (46) at (3, 4) {};
		\node [style=none] (47) at (3, 1.5) {};
		\node [style=none] (48) at (-1.5, 1.5) {};
		\node [style=none] (49) at (-1.5, -1.5) {};
		\node [style=none] (50) at (1, -1.5) {};
		\node [style=none] (51) at (1, -4.5) {};
		\node [style=none] (52) at (-1.5, -4.5) {};
		\node [style=none] (53) at (-0.75, 1.5) {};
		\node [style=none] (54) at (-0.75, -1.5) {};
		\node [style=none] (55) at (0.5, 1.5) {};
		\node [style=none] (56) at (0.5, -1.5) {};
		\node [style=none] (57) at (1.5, 1) {};
		\node [style=none] (58) at (1.5, 1.5) {};
	\end{pgfonlayer}
	\begin{pgfonlayer}{edgelayer}
		\draw [qWire] (1.center) to (0.center);
		\draw (2.center) to (5.center);
		\draw (5.center) to (4.center);
		\draw (4.center) to (3.center);
		\draw (3.center) to (2.center);
		\draw (7.center) to (10.center);
		\draw (10.center) to (9.center);
		\draw (9.center) to (8.center);
		\draw (8.center) to (7.center);
		\draw [cWire] (14.center) to (13.center);
		\draw [cWire, in=90, out=-90] (16.center) to (15.center);
		\draw [cWire, in=150, out=-90, looseness=1.25] (20.center) to (19);
		\draw [cWire] (19) to (33.center);
		\draw [cWire, in=-90, out=30] (19) to (34.center);
		\draw [cWire] (34.center) to (35.center);
		\draw [cWire, in=-90, out=90] (35.center) to (36.center);
		\draw [cWire] (36.center) to (40.center);
		\draw [cWire, in=-90, out=90, looseness=0.75] (40.center) to (42.center);
		\draw [thick gray dashed edge] (48.center)
			 to (47.center)
			 to (46.center)
			 to (45.center)
			 to cycle;
		\draw [thick gray dashed edge] (52.center)
			 to (51.center)
			 to (50.center)
			 to (49.center)
			 to cycle;
		\draw [thick gray dashed edge] (53.center) to (54.center);
		\draw [thick gray dashed edge] (55.center) to (56.center);
		\draw [cWire] (58.center) to (57.center);
	\end{pgfonlayer}
\end{tikzpicture}},
\eeq
\caption{The scenario with the lab notebook modeled as a physical system (denoted $X$) which is on the same footing as $S$.}
\label{PMScenario2}
\end{center}
\end{figure}

To respond to this, it is useful to first recast the lab notebook objection into the language of  GPT states and the operational identities that hold among them. 
Imagine that the choice of preparation is copied and viewed as a physical system  $X$ on equal footing with the system $S$, as shown in Figure~\ref{PMScenario2}.
System $X$ plays the role of the lab notebook; its value constitutes the classical record of which preparation was performed. We denote a state of knowledge wherein one has certainty that $X$ takes value $x$ by ${\bf \delta}_x$. 
The GPT states on the composite system $SX$, therefore, are given by
\begin{equation} \label{thesetofstates}
\{ {\bf s}_x\otimes {\bf \delta}_x\}_{x\in X}.
\end{equation} 
These states are all linearly independent as GPT vectors, since $ \sum_{x\in X} \gamma_x {\bf s}_x\otimes {\bf \delta}_x =0$ if and only if $\gamma_x=0$ for all $x\in X$.  
The lab notebook objection is then expressible as follows: if we consider the operational states on the system $S$, then they satisfy nontrivial linear dependence relations of the form of Eq.~\eqref{linconstr},
but if we include the lab notebook $X$ in our analysis, then the operational states of the system and notebook are those of Eq.~\eqref{thesetofstates}, which are {\em linearly independent}, and hence do not satisfy any nontrivial relation of the form of Eq.~\eqref{linconstr} . In other words, including the notebook, the sceptic claims, leads to there being no nontrivial operational identities among the states. 
It is well-known that one cannot prove the failure of noncontextuality in a prepare-measure scenario without making use of some nontrivial operational identities among the states~\cite{gencontext}.  Hence, the argument goes, one can always find a noncontextual model for the scenario, viewed as an experiment on $SX$. 

Thus, proponents of the lab notebook argument claim that one reaches different verdicts for the exact same experiment, depending on whether or not one includes system $X$ as a causal mediary in one's analysis.

However, this is not correct.
When this scenario is correctly analyzed as an experiment on 
$SX$ as depicted in Figure~\ref{PMScenario2}, one gets the same answer as in the original analysis---the experiment {\em is} a proof of nonclassicality, even when conceptualized in this way.

The mistake arises from the belief that the linear independence of the states $\{ {\bf s}_x\otimes {\bf \delta}_x\}_{x\in X}$ implies that there are  no operational identities among the GPT states on $SX$. 
As we noted in Section~\ref{opequivsub}, not all operational identities take the form of bare linear dependence relations.
(This realization came in part from discussions with Ana Bel\'en Sainz, Elie Wolfe, and Ravi Kunjwal.)
Indeed,  if the GPT states on $S$ satisfy the operational identities in Eq.~\eqref{opidonS}, then the GPT states on $SX$ 
satisfy the operational identities
\begin{equation} \label{mainopeq}
\forall j: \sum_{x\in X}  \alpha^{(j)}_x  \mathsf{tr}_X({\bf s}_x\otimes {\bf \delta}_x) = 0.
\end{equation}
On the basis of this operational identity, one can derive a noncontextuality inequality that is violated in this scenario---namely, the exact same inequality that one arrived at via the original analysis of the scenario (the one which did not treat $X$ as a causal mediary on par with $S$). 

Explicitly: an ontological model for the composite system $SX$ posits\footnote{The fact that we take the ontic state space to be a Cartesian product of the ontic state spaces for $S$ and for $X$ could be viewed as a consequence of diagram preservation~\cite{schmid2020structure}. It also follows immediately from the causal structure assumed in the lab notebook argument---that system $X$ is a system whose role is to encode perfect classical information about which preparation was performed. (In fact, one can moreover conclude from the causal structure that $\Lambda_X$ is isomorphic to the set of possible values of $X$, and that $\mu_x(\lambda_S,\lambda_X)= \mu_x(\lambda_S)\otimes\delta_{\lambda_X,x}$, but the argument does not need this specificity.) } an ontic state space $\Lambda_S \times \Lambda_X$ and represents each of the states ${\bf s}_x\otimes {\bf \delta}_x$ by some probability distribution $\mu_x(\lambda_S,\lambda_X)$. The constraints implied by generalized noncontextuality together with the operational equivalence in Eq.~\eqref{mainopeq} is
\begin{equation} \label{margopeq}
\forall j: \sum_{x \in X} \alpha^{(j)}_x \sum_{\lambda_X\in \Lambda_X}\mu_x(\lambda_S,\lambda_X)=0
\end{equation}
 where we have made use of the fact that ${\rm tr}_
X$ is represented in the ontological model by marginalization over $\Lambda
_X$. 
But $\sum_{\lambda_X\in \Lambda_X}\mu_x(\lambda_S,\lambda_X) = \mu_x(\lambda_S)$, where $\mu_x(\lambda_S)$ is the distribution representing ${\bf s}_x$, so that Eq.~\eqref{margopeq} is simply
\begin{equation} \label{margopeq2}
\forall j: \sum_{x \in X} \alpha^{(j)}_x \mu_x(\lambda_S)=0
\end{equation}
which is simply Eq.~\eqref{origopeq}, the constraint one obtained in the original analysis---which, by assumption, leads to a noncontextuality inequality that is violated by the observed statistics in the experiment.

In short,
{\em whether or not} one chooses to treat the lab notebook as a dynamical system, one reaches the same verdict: the experiment in question does not admit of a noncontextual explanation. This was missed by proponents of the lab notebook objection because the full scope of possible operational identities was not recognized. 

As a final clarifying remark, we note that the assumptions underlying the use of operational identities in noncontextuality arguments are exactly the same as the assumptions underlying the use of the Bloch sphere as a representation of a qubit. 
Consider the case of a single qubit, as represented by the Bloch ball. By definition, the points in the Bloch ball describe operational equivalence classes of preparation procedures, where each point contains all and only the information needed to  predict the statistics of all measurements on the qubit. 
  As a concrete example, the centre point of the Bloch ball represents many different ways to prepare the maximally mixed state, such as taking an equal mixture of $\ket{0}$ and $\ket{1}$ or an equal mixture of $\ket{+}$ and $\ket{-}$. In any real experiment where one prepares states of the qubit, there will exist some records of what preparation was performed on the qubit. (In our example, this would be a record of whether $Z$ or $X$ eigenstates were prepared in a given run.) 

If one chooses to represent the joint state of the qubit together with these records, the density matrices one so obtains will be linearly independent, and consequently will form a simplex rather than a Bloch ball. This in no way undermines the fact that the preparations of the qubit satisfy operational equivalences, nor does it undermine the validity of the Bloch representation. 
When one computes the operational states of the system alone, 
after tracing out the lab notebook, one recovers the Bloch ball. 

To make sense of the Bloch sphere representation---just as is needed to make sense of generalized noncontextuality---one must assume that one can meaningfully single out a specific degree of freedom, and perform measurements on it (and it alone). Quantum physicists
(both theorists and experimentalists) 
 know how to study single systems in isolation, and how to characterize the GPT governing some such system $S$; recall, for example, the discussion of theory-agnostic tomography in Section~\ref{gpttomography}. The existence of any number of records or copies of this information, or of details about how this information was obtained, is irrelevant to this fact. And once one has a characterization of the GPT governing the system of interest, determining whether the system is classically-explainable or not is simply a matter of 
  testing simplex embeddability on it (or deriving and testing noncontextuality inequalities for it).

\section{Other common objections}

We now reply to a number of other objections to the notion of generalized noncontextuality.

\subsection{The physical-mixtures objection} \label{physinfer}

Another common objection (which is close in spirit to the lab notebook objection) is that the existence of classical records about what procedure was implemented precludes the possibility of defining or physically implementing mixtures of different laboratory procedures. Only if this record is somehow fundamentally erased from existence, the argument goes, could one hope to have implemented a true mixture of procedures. 
  
Perhaps the simplest response to this objection is to note that some proofs of noncontextuality---for example, those using the simplex-embedding approach---make no explicit reference to mixtures of preparations (or indeed even to mixed states). Similarly, experimental tests of noncontextuality within this approach (e.g., using theory-agnostic tomography) do not require one to implement any particular mixtures of given states. One can determine whether a given theory (or experiment) is classical or nonclassical based solely on the geometry of the state and effect spaces. 

However, there are insights to be learned by providing a more thorough analysis of this objection. It arises from a misunderstanding regarding the notion of a GPT state vector (or of a density operator, in quantum theory). Indeed, this is the same misunderstanding that sometimes leads to the claim (discussed in Sec.~\ref{sec:LabNotebook}) that one must include the lab notebook $X$ as a physical system in one's analysis. A GPT state vector is an equivalence class of preparation procedures, where the equivalence relation is defined relative to all measurements {\em on a given system}. And one can define and experimentally characterize GPT state vectors, regardless of the existence of any number of records or copies of information pertaining to which laboratory procedures were used to generate them. (See also our comments at the end of Sec.~\ref{sec:LabNotebook}.) 

In addition, this objection misses the fact that mixtures appearing in noncontextuality arguments can be (and should be) viewed as inferential rather than physical~\cite{schmid2020unscrambling,Pirsa_omelette}. That is, they need only describe the knowledge of 
agents who are reasoning about the  system. 
One need not imagine a dice-rolling procedure implemented physically to justify the applicability of a probabilistic mixture. 
 Based on whatever actual procedures one happens to have implemented, one can always leverage classical probability theory to reason about any hypothetical ensemble of procedures, where each of the actual procedures appears with some particular relative frequency in the ensemble. 
  Mixed states need not arise in any other capacity in noncontextuality scenarios.

A final related confusion concerns the distinction between proper and improper mixtures.
(Recall that a proper mixture is defined as a state of classical uncertainty about what quantum state describes a given system, whereas an improper mixture is defined as the marginal of some entangled bipartite pure state.) 
It is sometimes suggested that for a given mixed state, noncontextuality arguments presume that it is realized as a proper mixture and that this is somehow problematic in the sense that noncontextuality arguments are silent about improper mixtures. But neither of these is the case: rather, the details of how one prepares a given mixed state are irrelevant because all such preparation procedures are operationally equivalent. In other words, only the set of GPT states is relevant for questions of noncontextual realizability, and whether a given GPT state is realized as a proper or improper mixture is simply part of the preparation context and hence irrelevant to the ontological representation.

\subsection{The device-dependence objection}\label{sec:MATLAB}

Another frequent challenge to the notion of generalized noncontextuality rests on the fact that a classical computer can simulate any given set of prepare-measure statistics, even statistics that are not realizable within a noncontextual ontological model.
Does the classical computer itself not then constitute a classical explanation of the statistics?

We will first give a direct answer to this question, and an example to illustrate it. We will then return to a deeper discussion of some key surrounding issues.

In short: whether or not a scenario is deemed noncontextual does not rest merely on the bare statistics, but also on the operational identities holding among the processes which generated those statistics. 
A classical computer simulation of an experiment  
 fails to reproduce the operational identities
 that hold in  the experiment, 
  and so does not constitute a  good classical explanation of the experimental statistics because it has failed to achieve a good explanation of the operational identities that are observed in the experiment.  
  
This is best illustrated by a simple example (which was constructed in collaboration with Ana Bel\'en Sainz, Elie Wolfe, and Ravi Kunjwal). 
 Consider an experiment with two binary inputs, the setting variables $X$ and $Y$, and two binary outputs, the outcome variables $A$ and $B$.
 Suppose the correlations between 
 $A$ and $B$ conditioned on 
 $X$ and $Y$,  denoted $P(AB|XY)$, achieve the maximum possible violation of a CHSH inequality.  That is, suppose that
  \begin{equation}\label{PRboxcorrs}
 P(AB|XY) =  \tfrac{1}{2}([00] + [11]) \delta_{XY,0} +  \tfrac{1}{2}([01] + [10]) \delta_{XY,1} 
 \end{equation} 
 where we have used the shorthand notation $[ab] := \delta_{A,a}\delta_{B,b}$. 
 This is easily recognizable as the input-output correlation associated to a Popescu-Rohrlich box~\cite{Popescu1994}. 
 
\begin{figure}[htbp!]
\begin{center}
(a)\ \ $ %
\begin{tikzpicture}
	\begin{pgfonlayer}{nodelayer}
		\node [style=none] (0) at (1, -0.25) {};
		\node [style=none] (1) at (1, 3.25) {};
		\node [style=none] (6) at (1, 4.25) {};
		\node [style=none] (7) at (0.5, 3.25) {};
		\node [style=none] (8) at (2.5, 3.25) {};
		\node [style=none] (9) at (2.5, 4.25) {};
		\node [style=none] (10) at (2, 4.25) {};
		\node [style=none] (11) at (2, 5) {};
		\node [style=none] (13) at (2, 3.25) {};
		\node [style=right label] (15) at (2, 5) {$B$};
		\node [style=right label] (16) at (2, 2.5) {$Y$};
		\node [style=right label] (17) at (1, 1.25) {$S$};
		\node [style=none] (19) at (1.5, 3.75) {$M'_{\rm PR}$};
		\node [style=none] (25) at (1.75, -2.75) {};
		\node [style=none] (39) at (0, -0.25) {};
		\node [style=none] (40) at (4.5, -0.25) {};
		\node [style=none] (41) at (4.5, -4.5) {};
		\node [style=none] (42) at (0, -4.5) {};
		\node [style=none] (44) at (0.75, -0.25) {};
		\node [style=none] (54) at (0.5, -2.75) {};
		\node [style=none] (55) at (3, -2.75) {};
		\node [style=none] (56) at (1.75, -4) {};
		\node [style=none] (57) at (1.75, -3.25) {$\mathbf{s}_{\rm PR}$};
		\node [style=none] (64) at (2, 2.5) {};
		\node [style=none] (65) at (2, 3) {};
		\node [style=none] (66) at (2.5, -2.75) {};
		\node [style=none] (67) at (2.5, -1.75) {};
		\node [style=none] (70) at (2.5, -0.75) {};
		\node [style=none] (71) at (2, -1.75) {};
		\node [style=none] (72) at (4, -1.75) {};
		\node [style=none] (73) at (4, -0.75) {};
		\node [style=none] (74) at (3.5, -0.75) {};
		\node [style=none] (75) at (3.5, 0.25) {};
		\node [style=none] (76) at (3.5, -1.75) {};
		\node [style=right label] (77) at (3.5, 0.25) {$A$};
		\node [style=right label] (78) at (3.5, -5) {$X$};
		\node [style=right label] (79) at (2.5, -2.25) {$S$};
		\node [style=none] (80) at (3, -1.25) {$M_{\rm PR}$};
		\node [style=none] (92) at (3.5, -5) {};
		\node [style=none] (93) at (3.5, -4.5) {};
		\node [style=none] (94) at (1, -2.75) {};
		\node [style=none] (95) at (1, -0.25) {};
		\node [style=none] (97) at (2.5, -2.75) {};
	\end{pgfonlayer}
	\begin{pgfonlayer}{edgelayer}
		\draw [qWire, in=90, out=-90] (1.center) to (0.center);
		\draw (6.center) to (9.center);
		\draw (9.center) to (8.center);
		\draw (8.center) to (7.center);
		\draw (7.center) to (6.center);
		\draw [cWire] (11.center) to (10.center);
		\draw [thick gray dashed edge] (42.center)
			 to (41.center)
			 to (40.center)
			 to (39.center)
			 to cycle;
		\draw (56.center) to (55.center);
		\draw (55.center) to (54.center);
		\draw (54.center) to (56.center);
		\draw [cWire, in=-90, out=90, looseness=1.25] (64.center) to (65.center);
		\draw [cWire, in=90, out=-90] (13.center) to (65.center);
		\draw [qWire] (67.center) to (66.center);
		\draw (70.center) to (73.center);
		\draw (73.center) to (72.center);
		\draw (72.center) to (71.center);
		\draw (71.center) to (70.center);
		\draw [cWire] (75.center) to (74.center);
		\draw [cWire, in=-90, out=90, looseness=1.25] (92.center) to (93.center);
		\draw [cWire, in=90, out=-90] (76.center) to (93.center);
		\draw [qWire, in=90, out=-90] (95.center) to (94.center);
	\end{pgfonlayer}
\end{tikzpicture}}$ \hspace{1cm} (b)\ \ $%
\begin{tikzpicture}
	\begin{pgfonlayer}{nodelayer}
		\node [style=none] (0) at (1.25, -1.5) {};
		\node [style=none] (1) at (1.25, 2.25) {};
		\node [style=none] (6) at (1.25, 4.25) {};
		\node [style=none] (7) at (0.75, 3.25) {};
		\node [style=none] (8) at (3.75, 3.25) {};
		\node [style=none] (9) at (3.25, 4.25) {};
		\node [style=none] (10) at (2.25, 4.25) {};
		\node [style=none] (11) at (2.25, 5.25) {};
		\node [style=none] (13) at (3.25, 3.25) {};
		\node [style=right label] (15) at (2.25, 5.25) {$B$};
		\node [style=right label] (16) at (3.25, 1.75) {$Y$};
		\node [style=none] (19) at (2.25, 3.75) {$\delta_{xy,a'\oplus b}$};
		\node [style=none] (25) at (2.25, -2.75) {};
		\node [style=none] (35) at (0.25, 4.75) {};
		\node [style=none] (36) at (4.25, 4.75) {};
		\node [style=none] (37) at (4.25, 2.25) {};
		\node [style=none] (38) at (0.25, 2.25) {};
		\node [style=none] (39) at (0.25, -1.5) {};
		\node [style=none] (40) at (5, -1.5) {};
		\node [style=none] (41) at (5, -4.5) {};
		\node [style=none] (42) at (0.25, -4.5) {};
		\node [style=none] (43) at (1, 2.25) {};
		\node [style=none] (44) at (1, -1.5) {};
		\node [style=none] (45) at (2.25, 2.25) {};
		\node [style=none] (46) at (2.25, -1.5) {};
		\node [style=none] (54) at (0.75, -2.75) {};
		\node [style=none] (55) at (3.75, -2.75) {};
		\node [style=none] (56) at (2.25, -4.25) {};
		\node [style=none] (57) at (2.375, -3.175) {$\frac{1}{2}\delta_{a,a'}$};
		\node [style=none] (64) at (3.25, 1.75) {};
		\node [style=none] (65) at (3.25, 2.25) {};
		\node [style=none] (66) at (3, -2.75) {};
		\node [style=none] (67) at (4.25, -1.5) {};
		\node [style=right label] (77) at (4.25, -1) {$A$};
		\node [style=right label] (78) at (4.25, -5) {$X$};
		\node [style=none] (92) at (4.25, -5) {};
		\node [style=none] (93) at (4.25, -4.5) {};
		\node [style=none] (94) at (1.5, -2.75) {};
		\node [style=none] (95) at (1.25, -1.5) {};
		\node [style=none] (97) at (3, -2.75) {};
		\node [style=right label] (98) at (1.25, 0) {$A'$};
		\node [style=right label] (99) at (2, 0) {$X$};
		\node [style=none] (100) at (2, -1.5) {};
		\node [style=none] (101) at (2, 2.25) {};
		\node [style=none] (102) at (1.25, 2.25) {};
		\node [style=none] (103) at (1.25, 3.25) {};
		\node [style=none] (104) at (4.25, -2.75) {};
		\node [style=none] (105) at (2, -1.5) {};
		\node [style=none] (106) at (4.25, -1.5) {};
		\node [style=none] (107) at (4.25, -1) {};
		\node [style=none] (108) at (2.25, 3.5) {};
		\node [style=none] (112) at (2, 2.25) {};
		\node [style=none] (113) at (2.25, 3.25) {};
	\end{pgfonlayer}
	\begin{pgfonlayer}{edgelayer}
		\draw [cWire, in=90, out=-90] (1.center) to (0.center);
		\draw (6.center) to (9.center);
		\draw (9.center) to (8.center);
		\draw (8.center) to (7.center);
		\draw (7.center) to (6.center);
		\draw [cWire] (11.center) to (10.center);
		\draw [thick gray dashed edge] (38.center)
			 to (37.center)
			 to (36.center)
			 to (35.center)
			 to cycle;
		\draw [thick gray dashed edge] (42.center)
			 to (41.center)
			 to (40.center)
			 to (39.center)
			 to cycle;
		\draw [thick gray dashed edge] (43.center) to (44.center);
		\draw [thick gray dashed edge] (45.center) to (46.center);
		\draw (56.center) to (55.center);
		\draw (55.center) to (54.center);
		\draw (54.center) to (56.center);
		\draw [cWire, in=-90, out=90, looseness=1.25] (64.center) to (65.center);
		\draw [cWire, in=90, out=-90] (13.center) to (65.center);
		\draw [cWire, in=90, out=-90] (67.center) to (66.center);
		\draw [cWire, in=-90, out=90, looseness=1.25] (92.center) to (93.center);
		\draw [cWire, in=90, out=-90] (95.center) to (94.center);
		\draw [cWire, in=-90, out=90, looseness=1.25] (100.center) to (101.center);
		\draw [cWire, in=90, out=-90] (103.center) to (102.center);
		\draw [cWire, in=90, out=-90] (105.center) to (104.center);
		\draw [cWire] (107.center) to (106.center);
		\draw [cWire, in=90, out=-90] (104.center) to (93.center);
		\draw [cWire, in=90, out=-90] (113.center) to (112.center);
	\end{pgfonlayer}
\end{tikzpicture}}$
\caption{  Two experimental prepare-measure scenarios that achieve the correlations $P(AB|XY)$ of Eq.~\eqref{PRboxcorrs}.  (a) The causal mediary is a nonclassical GPT system, in which case the correlations are evidence of nonclassicality, and (b) the causal mediary is a classical system, in which case the correlations are {\em not} evidence of nonclassicality.  The difference between experiments (a) and (b) is manifested in the operational identities that hold among the preparations of the system being transmitted. 
The specific GPT states and measurements in (a) live in the GPT known as Boxworld~\cite{GPT_Barrett} and we follow the notation given in Eq. (8) of Ref~\cite{cavalcanti2022post}. 
}
\label{twoways}
\end{center}
\end{figure}
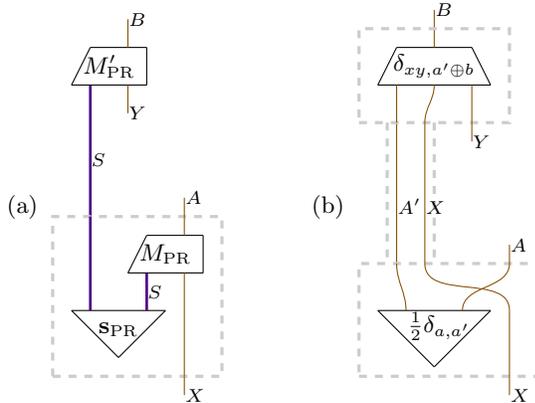

Now suppose that the experiment yielding these correlations is a bipartite Bell scenario, i.e., using measurements on a bipartite state. 
If the outcome at one wing is space-like from the mechanism choosing the value os the setting variable at the opposite wing, then observing the correlations in Eq.~\eqref{PRboxcorrs} implies that the experiment 
 cannot be explained by a locally causal ontological model.  Even if the measurements are not space-like separated, so that the experiment can be conceptualized as being of the prepare-measure variety, as depicted in  Fig.~\ref{twoways}(a), then observing the correlations in Eq.~\eqref{PRboxcorrs} implies that the experiment 
 cannot be explained by a noncontextual ontological model.
 In either of these scenarios, the correlations witness the impossibility of a certain type of classical explanation. 

Now suppose that  the experiment is again of the prepare-measure variety.  This time, however, suppose that  it is not a  nonclassical system  (i.e., a ``boxworld'' system)  that is transmitted from the preparation to the measurement, but a classical system that encodes  both $X$ and $A$, 
as depicted in Fig.~\ref{twoways}(b) (here $A'$ denotes a copy of $A$).  In this case, the experiment can still generate a conditional distribution $P(AB|XY)$  of the form of Eq.~\eqref{PRboxcorrs},  
 but such correlations no longer witness the failure of a noncontextual ontological model and hence no longer witness the impossibility of a classical explanation.
The reason that the realization of the correlations  of Eq.~\eqref{PRboxcorrs} in the context of the experiment of   Fig.~\ref{twoways}(a) exhibits nonclassicality but the realization of the {\em same} correlations  in the context of the experiment of  Fig.~\ref{twoways}(b) does not, 
  is because the two realizations
  satisfy {\em different operational identities}. In the former case, the two effective  GPT states on 
  system $S$  (one for each of the two possible values of $X$)   that arise when one marginalizes over $A$
   are  equal, 
   and it is this operational identity that allows one to derive the noncontextuality inequality that is violated by  the correlations of Eq.~\eqref{PRboxcorrs}.  
    In contrast, in the latter case, the two effective  GPT states of the causal mediary  $A'X$ are not the same (indeed, they are perfectly distinguishable, as $X$ takes a different value in the two states), 
    and so the noncontextuality inequality just mentioned is not a constraint on this scenario, and its violation cannot support any conclusions about noncontextuality.

We can now discuss some key surrounding issues.
The objection above is primarily raised by researchers who favour the device-independent paradigm for demonstrating quantum-over-classical advantages in information processing.  In a Bell scenario, the argument goes, one does not need to check any additional data to be sure that the observed statistics are nonclassical: one can check this from the observed correlations alone. It is often additionally claimed that this is a major advantage over tests of generalized noncontextuality.

However, it is not true that one does not need to check any additional data to be sure that the observed statistics are nonclassical in a Bell scenario. Rather, one must check that these statistics were generated in a particular causal structure: one where the outcomes are only connected by a common cause~\cite{wood2015lesson}. (This restriction on the classical simulation is typically formalized in terms of Bell's notion of locality causality.) As such, the mere presence of Bell inequality violations in an experimental scenario is not by itself sufficient to witness nonclassicality.
Similarly, the task of simulating noncontextuality inequality violations is only nontrivial if one takes into account additional empirical data: the operational identities among the operational states and effects  in the experimental set-up.

In Bell scenarios, this additional information (namely, whether or not the causal structure is one where the outcomes are only connected by a common cause) is often left implicit, which is why it is often said that one can decide if a given set of correlations is classical or nonclassical 
 simply by examining the correlations themselves. 
The reason this information is typically neglected is the belief that it is quite independent 
of the system whose nonclassicality is being probed. Indeed, the typical way to justify such a claim---that the causal structure is one where the outcomes are only connected by a common cause---is to appeal to the theory of relativity, together with the experimental evidence that the choice of setting on each wing is space-like separated from the outcomes at the other wing. This evidence comes in the form of distance and timing measurements, which are presumed to be quite independent from the measurements on the entangled quantum systems.

In current schemes for experimentally testing noncontextuality, 
by contrast, the additional information one gathers to assess classicality comes from additional preparations and measurements on the degree of freedom whose nonclassicality is being probed. In particular, one finds the best-fit GPT representations of the preparations and measurements (from which one can extract operational equivalences if one desires).

This apparent contrast might seem at first to vindicate the claim that contextuality tests have a different status than Bell tests.  However, we now seek to show that the contrast is largely  illusory and likely to diminish 
further as better tests of noncontextuality are devised.

 While it is true that {\em current} tests of noncontextuality require a preliminary step of finding the best-fit GPT representations of one's preparations and measurements,
 it is possible that a future test
 of noncontextuality might be found where all of the operational identities that are used can be justified on grounds that are distinct from the experimental statistics gathered for preparations and measurements on the system in question. This would parallel more closely the type of empirical evidence used to justify the applicability of local causality in a  Bell test that closes the locality loophole. 
Whether this possibility is realized is an important open question for researchers studying noncontextuality.\footnote{If the possibility is realized, then one could test noncontextuality without first obtaining the best-fit GPT representations of the preparations and measurements.  Rather, one could simply do a hypothesis test on the possibility of a noncontextual model by looking for representations of the preparations and measurements as classical distributions and response functions over the ontic state space respectively and demanding that these satisfy the constraints implied by the operational identities.  If no such representations can be found that yield a good fit to the data, one has ruled out the hypothesis of a noncontextual model. }
 
Furthermore, space-like separation of the wings of the experiment is not the only sort of evidence one can leverage to support a conclusion of nonclassicality in a Bell experiment.  In other words,  one can have strong evidence of nonclassicality in Bell experiments even when the locality loophole is {\em not} closed.  For example, the statistical data accumulated in the experiment can provide evidence for nonclassicality because the 
classical explanations involving cause-effect relations between the wings {\em overfit} the data relative to explanations involving a quantum common-cause~\cite{daley2021experimentally}.  The latter sort of demonstration of nonclassicality in a Bell scenario has a close analogue  in contextuality scenarios.  

 The claim that noncontextuality tests are different from Bell tests is also undermined by the fact that the widespread claim that 
nonclassicality in Bell tests can be inferred {\em from the observed correlations alone} is not accurate. 
Specifically, we argue that one cannot, strictly speaking, implement a Bell test by simply taking the finite-run relative frequencies seen in the experiment and plugging these into the left-hand side of a Bell inequality. 
 
All real-world experiments are finite-run and all finite-run statistics include fluctuations.  For this reason, 
no real-world experiment yields the true probabilities, i.e., the relative frequencies that would be observed in an idealized limit of infinitely many samples.  
Nonetheless, there are certain constraints on the true probabilities that an experimentalist might know to hold.  For instance, in a Bell test,  if an experimenter is confident that there is space-like separation between the wings 
and they are confident in the correctness of relativity theory, then they can assume, as a constraint on their estimate of the true probabilities, that it must satisfy the no-signalling condition.   The finite-run relative frequencies, however, will generally {\em violate} the no-signalling condition simply because of statistical fluctuations.\footnote{ Specifically, for any finite-run statistics, the relative frequencies of outcome values at Alice's wing will generally show slight differences for  different settings values at Bob's wing simply because of statistical fluctuations. } 
 Therefore, to find an estimate of the true probabilities that respects the no-signalling condition, one cannot use the na\"{i}ve procedure of taking the relative frequencies as estimates of the true probabilities. 

A more methodologically sound analysis technique for experimental tests of Bell's notion of local causality (in the sense of the methodology of statistical model selection) 
estimates the true probabilities through a fitting procedure.  For instance, one can adopt a statistical model for the hidden variable source (while  assuming, without loss of generality, that  the measurements respond deterministically in a prescribed manner~\cite{Fine}) and one can implement an optimization algorithm to find the best-fitting such model where the quality of fit is given by a measure of distance between the probability distribution that is predicted by the model and the relative frequencies that are observed in the experiment.  Such an analysis of a Bell experiment was implemented in Ref.~\cite{daley2021experimentally}  and used to rule out a locally causal model via a hypothesis test (though this was not the focus of that article).  Similar techniques for implementing a hypothesis test of local causality have been used to contend with the memory loophole~\cite{barrett2002quantum}  in Bell tests, as described in Refs.~\cite{bierhorst2015robust,shalm2015strong}.  Such techniques use the raw frequencies to find an estimate of the true probabilities while satisfying certain constraints, and then evaluate the Bell inequalities on these best-fit probabilities rather than on the raw relative frequencies.\footnote{ In short, {\em both} Bell tests and noncontextuality tests must engage in finding best-fit {\em classical} representations of preparations and measurements satisfying certain constraints.  This fitting procedure is distinct from the one that arises in current tests of noncontextuality wherein one must find best-fit {\em GPT} representations of the preparations and measurements.  This step is what defines the constraints that the classical fit must satisfy.  Although the GPT-fitting step is currently unique to tests of noncontextuality, it may be possible to circumvent it, as we noted above.} This undermines the claim that such fitting procedures are unique to tests of noncontextuality and hence also the claim that they constitute a way in which noncontextuality tests are different in kind from Bell tests. 

 Finally, we dispute the claim that Bell tests and noncontextuality tests are contrasting because the former are theory-independent while the latter are not. In fact, tests of generalized noncontextuality, just like Bell tests, do not need to make any prior assumption of the correctness of quantum theory,  nor do they need to make any prior assumption about the identity of each state or measurement used in the experiment. This is most evident in tests of generalized noncontextuality based on theory-agnostic tomography, as discussed in Sec.~\ref{gpttomography}. In these tests, one extracts from the data (rather than assumes) both the  dimension of the GPT vector space needed to model the system and the precise characterizations of the GPT state vectors and effect vectors that best fit the realized preparations and measurements.

Indeed, when one  takes the trouble to implement the greatest possible diversity of  laboratory procedures on a given system, theory-agnostic tomography can provide evidence for tomographic completeness of the realized set (i.e., that the realized GPT state and effect vectors span the true state and effect spaces respectively)  in the sense that one has the opportunity in such an experiment to {\em falsify} the hypothesis that some set of procedures are tomographically complete.  It is this possibility of falsification that makes it clear that one is not merely {\em assuming} tomographic completeness but gathering evidence for it.  
   In such a case, one can argue that evidence of nonclassicality can be reached directly from the observed data alone, with no extra assumptions. Whether or not this evidence is compelling depends on the extent to which the experiment really had an opportunity to falsify the hypothesis and hence on  one's confidence that the laboratory procedures in the experiment do in fact span the true state and effect space of the system being probed (or a valid GPT subsystem thereof, in the sense defined in Ref.~\cite{NCsubsystems}).  This point is discussed at greater length in the introduction of Ref.~\cite{mazurek2017experimentally}.
   
  In our view, the possibility that the procedures one has experimentally implemented fail to span the true state and effect space of the system or subsystem being probed is the most significant loophole for tests of noncontextuality.  No matter how hard one has tried and failed to falsify the hypothesis of tomographic completeness of a given set of procedures, it may be that at some future date, a novel experiment succeeds in achieving the falsification.  
  But there is parallel kind of loophole in a Bell test.  The claim that the two laboratories in a Bell experiment are space-like separated is also one that is based on empirical data, and no matter how much data one has accumulated in favour of this assessment, it might be falsified. This is clear if one thinks about the problem of verifying space-like separation as a two-party cryptographic task in the presence of an adversary.\footnote{ Suppose Alice and Bob seek to confirm that given events in their laboratories are indeed space-like separated, while an adversary seeks to fool them into thinking these events are space-like separated when in fact they are not.  Suppose, for instance, that Alice and Bob try to synchronize their clocks by a procedure wherein they transfer light signals to one another.  If the adversary adds delays to these signals, he can cause Alice and Bob to have false beliefs about what clock readings correspond to synchronization, and hence false beliefs about what events are space-like separated.  Similarly, whatever protocol Alice and Bob use for seeking to estimate the distance between their laboratories, the adversary could seek to interfere with that protocol as well.  It would be interesting to try and devise a protocol that could provide a guarantee of space-like separation (relative to some set of background assumptions) in the presence of an adversary. As far as we aware, no proposals for such a protocol have been made to date.}

Both Bell tests and contextuality tests are also theory-laden in another sense. 
Imagine that one is seeking to establishing space-like separation of a pair of events that are separated by a distance $d$.  This requires that one has timing precision of order $d/c$ where $c$ is the speed of light.  A skeptic may then wonder on what grounds one is confident that one's clock in fact has this kind of precision.  Generally, the grounds for such confidence always refer to our understanding of how the clock works according to our best physical theories. 

The point is this: the sort of evidence one can have for characterizing the causal structure and representation of experimental procedures in a noncontextuality test is not dramatically different in kind from the sort of evidence one can have for characterizing the causal structure and representation of experimental procedures in a Bell test.

\subsection{The efficient-simulability objection}

In certain circles, it is common to assume the following desideratum for a good notion of classical-explainability: that a given computational process (for instance, a quantum computation) should count as classically explainable if and only if it can be efficiently simulated on a classical computer. 
 But it is well known that there are subtheories of quantum mechanics that are efficiently simulable on a classical computer but that still exhibit contextuality. 
 For example, the stabilizer subtheory for qubits is efficiently simulable 
  due to the Gottesman-Knill theorem~\cite{gottesman1998heisenberg}, and yet is contextual, due to the possibility of realizing the GHZ or Mermin-square proofs of contextuality within it. (The result can be generalized to stabilizer subtheories in any even dimension~\cite{Schmid2022Stabilizer}.)
Consequently those who endorse the desideratum see this as a deficiency of generalized noncontextuality as a notion of classical explainability.

However, the idea that a notion of classical explainability must reproduce the divide between efficient and inefficient classical simulability is, in our view, unmotivated.  Quantum computation forms only a small subset of the scope of all physical phenomena, and there is no reason to expect that every manifestation of nonclassicality must be useful for the specific task of universal computation. 

For example, consider the kind of nonclassicality arising in Bell scenarios (which we take to be nonclassicality of the common cause~\cite{Wolfe2020quantifyingbell,Schmid2020typeindependent,Schmid2023understanding}). 
It is generally thought that this is a meaningful and interesting notion of nonclassicality. And yet, as noted in the previous section, it is only when one imposes constraints on the simulation (specifically, a constraint on its causal structure) that there is any challenge to simulating Bell inequality violations on a classical computer. 
In the case of a prepare-measure scenario, if one adopts generalized noncontextuality as one's notion of classical explainability, then the question of interest is whether one can simulate the experiment while 
respecting specific identities on the classical representations of states and specific identities on the classical representation of effects, namely, those that mirror the identities that hold among the states and effects themselves. 
 
 Different notions of classical explainability,
 we believe, correspond to different assumptions about what {\em constraints} a classical model of some operational phenomena ought to satisfy.  To evaluate the merit of a given notion is to evaluate the motivations for the constraints it proposes.  Conceiving of the classical model as a classical simulation, in the sense of computational complexity theory, does not alleviate the need to make such an assessment. 
For instance, there are many different computational complexity classes for which one can define a classical and a quantum version.  What differs between these classes is {\em what constraints} are imposed, for instance, the spatial and temporal resources that the computation is permitted to use.  

As an example, Anders and Browne~\cite{browne} consider a model of computation that is a version of measurement-based quantum computation, but where the classical processor which acts on the setting and outcome variables of the measurements can only make use of gates whose Boolean output is a {\em linear} function of the Boolean inputs (e.g., it can implement {\tt XOR} and {\tt NOT} gates, but it cannot implement an {\tt AND} gate).  They showed that in this model of computation, if one supplements the classical linear processor with a bipartite state and local measurements that are able to achieve the algebraically maximal violation of the CHSH inequalities (i.e., a Popescu-Rohrlich (PR) box~\cite{Popescu1994}), then one can implement an {\tt AND} gate on the Boolean inputs to the circuit.  The PR box correlations have promoted the computational power of this model from the parity-L class to universal classical computation.  
Because there are ways of implementing {\tt AND} gates that do not require access to Bell-inequality-violating correlations, the nonclassicality of the state and measurement resources is not witnessed by the ability of the circuit to go beyond universal classical computation, but rather by its ability to go beyond the parity-L class.  What this example suggests is that a given computational architecture can be judged to witness nonclassicality if the operational statistics it generates cannot be explained by a classical model {\em that respects the causal structure of that architecture}.  This may happen even though the computational task it achieves (such as implementing an {\tt AND} gate) is only difficult to achieve classically {\em relative to this causal structure}.  This example has been discussed in greater detail in Ref.~\cite{RobPIRSA}.   It has also been shown that in a measurement-based model of computation where the classical processor is linear, the power of the model can be increased by correlations that exhibit {\em contextuality}~\cite{comp1}.

It is also worth noting that, {\em a priori}, there is no reason to think that a notion of nonclassicality that was entirely motivated by questions about {\em computational complexity} would be able to explain advantages for other information-processing tasks, such as communication and cryptography, in particular, the known advantage that generalized contextuality implies for certain types of random access codes~\cite{RAC,RAC2}.

\subsection{The parochial-equivalences objection}

Another objection one often hears is that whether a pair of preparations are deemed to be operationally equivalent or not depends on what measurements one has made in a given experiment, or on what measurements can be made using current technology. If this were true, it would completely undermine generalized noncontextuality as a foundational notion of nonclassicality, since verdicts of classicality would be determined more by current technology and choices of what experiments to carry out than it would be by fundamental physics.

However, the notion of operational equivalence for preparations on a system, as defined in Ref.~\cite{gencontext}, is equivalence of statistics for all measurements that are possible {\em by the lights of the operational theory one is assuming.}

  Since a {\em tomographically complete} set of measurements is one such that its statistics are sufficient to infer the statistics of any other measurement, one can define operational equivalence of preparations on a system in terms of equivalence of the statistics of {\em all} measurements in a set that is tomographically complete by the lights of the operational theory. 
In short, operational equivalence is a notion that is only defined relative to an operational theory. 
If one assumes the correctness of quantum theory, then whether two preparations are operationally equivalent or not is assessed relative to a set of measurements that is tomographically complete {\em by the lights of quantum theory}.  
By contrast, if one assumes that a system is governed by some other GPT, distinct from quantum theory, then operational equivalences of preparations must be assessed   relative to a set of measurements that is tomographically complete by the lights of that GPT.

It is helpful to consider a thermodynamic example that is sometimes put forward to elucidate the objection, and to see in what way it
  misunderstands the definition of operational equivalence.

Consider an ideal gas of particles assumed to be governed by classical Newtonian mechanics, and consider a box with two compartments separated by a divider. Let us now define two different preparation procedures on the gas. For the first preparation, the gas is prepared at a specified temperature and pressure and such that it lies entirely in 
 the left compartment (while the right compartment is empty); then, the divider is removed so that the gas expands into the entire box.
 The second preparation procedure is identical, but where the gas begins in the right compartment (while the left is empty) prior to  removing the barrier. 
 We have thereby described two distinct preparation procedures,  in each of which the gas ends up 
   distributed throughout the whole box.   
These two preparation procedures lead to the same macroscopic thermodynamic properties (in particular, temperature and pressure) for the gas, but they correspond to different microstates. (This follows from the reversibility of Newtonian dynamics and the fact that the microstates at the initial time are different).
It follows that the two preparations are indistinguishable by any measurement of macroscopic thermodynamic properties (and perhaps even indistinguishable by any practically realizable measurement given current technology), but they are nonetheless associated to different ontological states.  

Therefore, {\em if} this type of indistinguishability of preparations was sufficient to infer their operational equivalence, then the two preparations would be operationally equivalent but represented by distinct distributions over the ontic states, and hence we would have described an example of preparation contextuality.   If this were the case, 
then the example  would undermine  the notion of generalized noncontextuality insofar as contextuality is being proposed as a notion of nonclassicality and a system governed by Newtonian mechanics ought not to be assessed as nonclassical. 

But the type of indistinguishability described here is not sufficient for inferring operational equivalence, and so the thought experiment does not constitute an example of preparation contextuality. 

In other words, the thought experiment only gives the appearance of undermining the notion of noncontextuality {\em if} one forgets that operational equivalences are defined relative to the set of all measurements that are possible in principle {\em by the lights of the physical theory one is assuming}. In Newtonian mechanics, the pair of preparations in the thought experiment are not, in fact, operationally equivalent.   This is because, by the lights of Newtonian mechanics, there is nothing forbidding a measurement that determines the exact microstate of the gas---the positions and momenta of each individual particle in the gas. One could then uniquely determine whether the microstate  at the final time 
 arose from 
  time evolution of  a microstate at the initial time wherein the gas started in  the left compartment or from one wherein the gas started in the right compartment. 
Such a measurement is obviously an incredible technical challenge, but it is not ruled out by the lights of the physical theory being assumed.
Indistinguishability relative to macroscopic thermodynamic properties or relative to measurements that are technologically feasible at the present day is {\em simply not relevant} to assessments of operational equivalence. 
 Rather, all that matters is operational equivalence relative to  the measurements that are possible in principle by the lights of the physical theory under consideration. 

In short, the objection considered here---that the notion of generalized noncontextuality  is undermined by the fact that technological capabilities dictate whether  in practice  two laboratory procedures are distinguishable or not---simply misunderstands the notion of operational equivalence.  For further discussion of this point, see Ref.~\cite{Leibniz} and Sec. II of Ref.~\cite{catani2022reply}.

Of course, if one wishes to assess directly from experimental data whether {\em nature} admits of a noncontextual model or not, then one cannot assume the correctness of any particular operational theory.  

One way around this problem is to seek to experimentally determine the set of operational theories that are consistent with the experimental data, using theory-agnostic tomography, and then to assess the possibility of a noncontextual model relative to this set of operational theories.  This is the approach taken in Ref.~\cite{mazurek2017experimentally}.

Theory-agnostic tomography requires that the set of preparations and the set measurements that are implemented on a system are {\em tomographically complete}. (Recall that a tomographically complete set of preparations is one such that its statistics are sufficient to infer the statistics of any other preparation, and a tomographically complete set of measurements is one such that its statistics are sufficient to infer the statistics of any other measurement.) But without prior knowledge of the operational theory governing a system, there is no way to know a priori whether the set of procedures that has been implemented is, in fact, tomographically complete.  This is not, however, a deficiency in the definition of an operational theory.  Rather, it is simply indicative of the fact that assessments of tomographic completeness, and hence of the operational theory that governs some system, are fallible.   This in turn implies that assessments of noncontextuality are also fallible.   This point is discussed further in Sec.~\ref{sec:fallibility}.

There is a second way to try and directly test noncontextuality without presuming the correctness of any particular operational theory. Given that operational equivalence is indistinguishability relative to all measurements and given that a tomographically complete set of measurements is, by definition, one such that indistinguishability relative to it implies indistinguishability relative to all measurements, it follows that to assess whether two preparations are operationally equivalent, it is sufficient to assess whether they give the same statistics for all measurements in a {\em tomographically complete set}.  Similarly, one can assess the operational equivalence of two measurements relative to a tomographically complete set of preparations. 
Thus, one can simply seek to assess operational equivalences in an experiment relative to a tomographically complete set of procedures (without seeking to meet the higher bar of determining all of the details about the GPT governing the system). Of course, assessments of tomographic completeness are fallible, but this merely implies that assessments of noncontextuality are also fallible, as we already noted above.

\subsection{The imperfect-equivalences objection}

Another concern which is sometimes raised about tests of generalized noncontextuality is that it is unclear how to
ensure that any given operational equivalence holds {\em exactly} between the procedures in any real experiment. 

Imagine that one is interested in a particular operational identity between some target states $\{ {\bf s}_x\}_{x\in\{1,2,3,4\}}$, say $\frac12 {\bf s}_1  + \frac12{\bf s}_2 = \frac12 {\bf s}_3 + \frac12 {\bf s}_4$, and imagine that one has derived a noncontextuality inequality from this operational identity. In a real experiment, one can never succeed at preparing any of these target states exactly, but rather one generally ends up preparing some alternative nearby state,  
which we denote $\{ \overline{\bf s}_x\}_{x=1,2,3,4}$.
The latter states will generally not satisfy  the operational identity $\frac12 \overline{\bf s}_1  + \frac12\overline{\bf s}_2 = \frac12 \overline{\bf s}_3 + \frac12 \overline{\bf s}_4$ that one was targeting. Consequently, the noncontextuality inequality one wished to test is strictly not relevant to the experiment one actually performed, and it seems one is blocked from ever getting noise-robust tests of noncontextuality.

However, there are (at least) three different ways of circumventing this problem. 

The first way to avoid this problem was introduced in Ref.~\cite{robust,mazurek2016experimental}. Basically, the proposed resolution is to recognize that if one has experimentally determined the operational statistics generated by some set of GPT states (or GPT effects), then one can {\em logically infer} the statistics that would be generated by any convex mixture of these.  So, one simply identifies a set of so-called {\em secondary states} and {\em secondary effects} that lie within the convex hull of those that were actually realized in the experiment, and that moreover satisfy exactly the desired operational identities. Although these states (effects) do not characterize any of the procedures that were actually implemented, they are known to be part of the operational theory governing the experiment, as they correspond to mixtures of procedures that were in fact realized and every operational theory is closed  under mixing.
One then tests the noncontextuality inequalities on the statistics described by these secondary states and effects (which can easily be computed from the states and effects).  If one finds that the inequalities are violated, then one can be certain that there is no noncontextual model of the experimental data.
 The downside of this approach (noted in Ref.~\cite{mazurek2016experimental}) is that although one can always find secondary states and effects that satisfy the desired operational equivalences, these are always noisier than the realized ones.  Since every noise-robust noncontextuality inequality has a threshold of noise beyond which it cannot be violated, it can happen that the transition from the primary to the secondary states and effects adds sufficient noise that one crosses the threshold and is unable to violate any nonconextuality inequality. 
  
The second approach is more direct, and does not require introducing any secondary states or effects.
Rather than deciding beforehand which noncontextual inequality is to be tested and consequently which operational identities are to be targeted in the experiment,  instead one simply characterizes the GPT states and GPT effects that are actually realized in the experiment, then one determines the operational identities that happen to hold among these, and one derives noncontextuality inequalities based on {\em these} operational identities.

In our example above, for instance, the four realized states were denoted $ \overline{\bf s}_1$, $ \overline{\bf s}_2$, $ \overline{\bf s}_3$ and $ \overline{\bf s}_4$, and it was noted that they in general will not satisfy the simple operational identity 
$\frac12 \overline{\bf s}_1  + \frac12\overline{\bf s}_2 = \frac12 \overline{\bf s}_3 + \frac12 \overline{\bf s}_4$.  
Nonetheless, if these states are confined to a two-dimensional state space\footnote{If the states are not confined to a two-dimensional state space, then one simply requires more than four states to have a nontrivial linear dependence relation and hence a nontrivial operational identity.},
then there will always exist 
real values $\{\alpha_x\}_{x\in\{1,2,3,4\}}$ for which $\alpha_1 \overline{\bf s}_1  + \alpha_2\overline{\bf s}_2 = \alpha_3 \overline{\bf s}_3 + \alpha_4 \overline{\bf s}_4$, and these values can be inferred from the experimental characterization of the states. It then suffices to determine what noncontextuality inequalities follow from this operational identity.
As it turns out, computing the noncontextuality inequalities that follow from an arbitrary set of operational identities can be achieved using a linear program~\cite{Schmid2018}.

The final approach circumvents the direct consideration of operational identities and noncontextuality inequalities altogether. One simply follows the procedure of theory-agnostic tomography (outlined in Sec.~\ref{gpttomography} and discussed in detail in Refs.~\cite{mazurek2017experimentally,grabowecky2021experimentally}) to experimentally determine the set of GPTs that are consistent with the experimental data. One then tests whether all of these GPTs are simplex-embeddable. Testing for simplex-embeddability is also achievable using a linear program~\cite{selby2024linear}. 

\subsection{The Kochen-Specker-were-na\"{i}ve objection}

Another objection that we have heard (in particular, from philosophers of physics) is that assumptions of noncontextuality are na\"{i}ve and unmotivated, and are studied today only because Kochen and Specker oversold their eponymous theorem. Recall that both Kochen and Specker~\cite{KS} and Bell~\cite{Bell2} independently arrived at no-go theorems from an assumption of noncontextuality (which we will here term {\em KS-noncontextuality}, in order to distinguish it from the assumption of generalized noncontextuality).
 However---the argument goes---Kochen and Specker did not emphasize the role of this assumption when summarizing their no-go result, stating simply that~\cite{KS}:
 {\em ``The main aim of this paper is to give a proof of the nonexistence of hidden variables.'' }
 Bell, by contrast, was more circumspect~\cite{Bell2}:
\begin{quote}
That so much follows from such apparently innocent assumptions leads us to question their innocence. Are the requirements imposed, which are satisfied by quantum mechanical states, reasonable requirements on the dispersion free states? Indeed they are not [...] It was tacitly assumed that measurement of an observable must yield the same value independently of what other measurements may be made simultaneously [...] These different possibilities require different experimental arrangements; there is no {\em a priori} reason to believe that the results [...] should be the same. The result of an observation may reasonably depend not only on the state of the system (including hidden variables) but also on the complete disposition of the apparatus.
\end{quote}
Indeed, it is well-known that 
one can construct explicit hidden variable models that do not satisfy the assumption of KS-noncontextuality 
 and that reproduce all of quantum theory---Bohmian mechanics is one example. 
The critics of noncontextuality, particularly those who find Bohmian mechanics to be a satisfactory interpretation, take this fact as evidence that Kochen and Specker's endorsement of the assumption as a natural one 
was na\"{i}ve and that we   should reject the assumption of KS-noncontextuality as unreasonable, as Bell suggests in the above quote. 

First, let us note that this is a rather uncharitable reading of Kochen and Specker's work, 
as they certainly recognize the possibility of hidden variable theories that violate their assumption. Just prior to the quote that is cited by their critics, for instance, they state:
\begin{quote}
There are on the one hand purported proofs of the non-existence of hidden variables, most notably von Neumann's proof, and on the other, various attempts to introduce hidden variables such as de Broglie~\cite{de1960non} and Bohm~\cite{Bohm} and \cite{Bohm2}. One of the difficulties in evaluating these contradictory results is that no exact mathematical criterion is given to enable one to judge the degree of success of these proposals.
\end{quote}

Nonetheless, it seems to us fair to say that Kochen and Specker did not articulate any clear a priori motivation for their assumption of noncontextuality. Bell, by contrast, stated outright that he did not see any good argument in favor of such a principle of noncontextuality.

While it may be true that no good argument in favour of endorsing KS-noncontextuality had been given at the time of Bell and Kochen-Specker's writings, such an argument
 {\em was}  provided in subsequent work: 
 one can motivate noncontextuality using a methodological  principle for theory construction due to Leibniz (a version of his principle of the identity of indiscernibles) 
  that has a long history of success in physics~\cite{Leibniz}. This principle motivates both KS-noncontextuality and also the notion of generalized noncontextuality introduced in Ref.~\cite{gencontext}.

Moreover, we consider the proof that one can characterize noncontextuality  as simplex-embeddability within the framework of GPTs~\cite{SchmidGPT} to constitute another motivation for taking it as a good notion of classical explainability.
For any simplex-embeddable GPT system, all the statistics that can be observed are compatible with the hypothesis that a {\em strictly classical} GPT gives the true description of one's system. 
This is because one can never establish by empirical means that an apparent restriction on states and effects---i.e., a restriction to a state space that is a strict subset of the full simplex and/or to an effect space that is a strict subset of the full hypercube of effects---is fundamental as opposed to merely being due to a technological limitation that might be overcome in the future.\footnote{A concrete example helps to illustrate the point.  If one performs theory-agnostic tomography on a system, and the state and effect spaces one realizes in the experiment are found to approximate those of the stabilizer states and measurements for a qubit (which are also the states and effects of the simplest system in the toy theory of Ref.~\cite{spekkens2007evidence}), then the range of GPTs that are consistent with this experimental data includes the strictly classical (i.e., simplicial) GPT of dimension 4.}
 In short, any experimental data that can be realized by a simplex-embeddable GPT 
can also be realized by a strictly classical GPT.   But strictly classical GPTs have been motivated~\cite{GPT_Barrett} (independently of any Leibnizian arguments) to be {\em the} GPT description of a system which is classical in the usual sense of being describable by a set of random variables (the different valuations of which define the possible ontic states of the system) which can be measured perfectly. 
 So simplex-embeddability is a natural notion of classical explainability.
 
Another set of motivations (which is less precise but arguably as compelling as those above) arises from inspection of the epistemically restricted classical theories in Refs.~\cite{spekkens2007evidence,bartlett2012reconstruction,Catani2023whyinterference,epistricted}. 
These theories are noncontextual and provide a compelling explanation of the operational phenomena that they reproduce.  But more than this, it is the noncontextuality of the theories that {\em makes} the explanations compelling, and it is for this reason that we take such theories to provide further evidence of the naturalness of the principle of generalized noncontextuality.
 For example, a distinctive feature of quantum theory is that a given mixed state can be convexly decomposed into an ensemble of pure states in many different ways.
Ontological models that are preparation-noncontextual explain the multiplicity of these different convex decompositions by modelling pure quantum states as non-point distributions over the ontic state space, and using the fact that many different mixtures of non-point distributions may yield the same distribution. (See, e.g., Sec.~III.A.4 of Ref.~\cite{spekkens2007evidence}.) Thus, noncontextuality provides a natural explanation of the multiplicity of convex decompositions of a mixed state in those subtheories of quantum theory that admit of a noncontextual ontological model.

For other motivations for taking generalized noncontextuality as a notion of classicality,
  see Refs.~\cite{Leibniz,schmid2020unscrambling}, or the introductions of Refs.~\cite{schmid2021guiding,selby2024linear}.

In any case, if sceptics wish to criticize the a priori naturalness of assumptions of noncontextuality, it is obviously insufficient for them to base their
 criticisms only on one or two writings that are half a century old.  
 They must also engage with all of the more recent motivations just discussed. 

\section{Further Discussion}

In the following sections, we expand on some of the above points, or discuss related ideas that we think deserve wider recognition. 

\subsection{Fallibility of assessments of contextuality}\label{sec:fallibility}

If one assumes the correctness of some particular operational theory (e.g., quantum theory), then the question of whether the theory (or a fragment of it) admits of a noncontextual ontological model can be settled by a theoretical investigation. In particular, one can derive the relevant operational identities from the theory, and derive a no-go theorem based on these. No experiment needs to be performed in this case.

Consider now the question of whether a given set of {\em experimental procedures} that are realized in the lab admit of a noncontextual model. Here, one may or may not wish to assume the correctness of some particular operational theory, but one does {\em not} presume to know how each laboratory procedure is represented in the theory. If one does assume the correctness of, say, quantum theory, then the question one is answering is whether or not one's experiment lives inside a fragment of quantum theory that is classically explainable, or whether one has accessed a broad enough fragment of quantum theory to be provably nonclassical. (This can be useful for the purposes of benchmarking experimental procedures that one hopes to use in a quantum information-processing task.) The highest bar, however, is to test whether {\em nature itself} is noncontextual. To do this, one cannot assume a priori the correctness of any particular operational theory.

In either of these last two cases, one must deduce the characterizations of one's laboratory procedures from experimental data.  
Typically, one focusses on a particular type of system, and one considers a  prepare-measure experiment on it.  Assessments of the possibility of a noncontextual model for this experiment are based on the operational identities that are found therein, or equivalently, on the shape of the fragment of the space of GPT states and GPT effects that are realized in the experiment.
Of course, if one is mistaken about the latter, then this will lead to mistaken conclusions about noncontextual-realizability and thus classical explainability.
As with any inference from finite-run data to a scientific hypothesis, 
the inference one makes from the data of a contextuality experiment to the characterization of the GPT states and effects (and hence the operational identities among these) might be mistaken.
 However, one {\em can} build up evidence for or against a given hypothesis about operational identities.
This evidence can be empirical, for instance, based on the best-fit states and effects arising from theory-agnostic tomography.  But it can also come from physical principles, such as locality or the absence of retrocausation.  
 It might also come from appealing to a particular physical theory (and the full body of evidence one has in support of that theory) and our knowledge of how that theory is applied to describe the particular laboratory procedures in question.

When the evidence in favour of operational identities among GPT states or among GPT effects comes from empirical data, the main way in which such assessments might ultimately prove to be incorrect is if the experimenter is mistaken about what constitutes a tomographically complete set of procedures. We refer the reader to Refs.~\cite{mazurek2017experimentally,grabowecky2021experimentally} for a discussion of this issue. It was also discussed at the end of Sec.~\ref{sec:MATLAB}.

It also  important, however, to study precisely when 
 mistaken assumptions about operational identities (i.e., the shapes of fragments of the GPT state and effect spaces) lead to mistaken assessments of noncontextuality. 

A first important result in this vein was given in Ref.~\cite{PuseydelRio}, which gave some sufficient conditions under which one can prove the failure of noncontextuality {\em even in cases where one is mistaken about or unsure of the operational identities}. A second important result in this vein follows from Lemma~11 of Ref.~\cite{muller2023testing}, concerning the question of whether contextuality proofs that assume the correctness of quantum theory are still valid if the world is in fact described by a postquantum GPT. In particular, the authors prove that such proofs continue to hold in the postquantum theory under some very reasonable assumptions about how quantum theory emerges from the postquantum theory via a decoherence-like process.
An elaboration on this result and related matters is provided in forthcoming work~\cite{NCsubsystems}. For example, Ref.~\cite{NCsubsystems} shows that robust proofs of contextuality are possible using only a subsystem or a subspace of a larger physical system. So, for example, the mere existence of unprobed degrees of freedom---internal or otherwise---do not in and of themselves undermine contextuality proofs.

\subsection{Noncontextuality is evaluated relative to a causal structure} \label{NCwrtcausalstr}

In general, assessments of noncontextuality can only be made relative to a causal structure. 
This has been obscured by the fact that most research to date has focused on the simplest case of prepare-measure scenarios, where the causal structure was too simple to have merited any discussion.

Even in this simplest case, however, one must make assumptions of a causal nature---for example, that the experiment can be conceptualized as a preparation of a system followed by a measurement of that system, and that this system acts as the complete causal mediary between the two stages.
This is the system relative to which one evaluates operational identities: e.g., two preparations of a system are deemed operationally equivalent if they give the same statistics for all measurements {\em on that system}. In other words, the system delimits the scope of the universal quantifier in the definition of operational equivalence.
The notion of system here is deemed to be a primitive notion, just as it is in the framework of GPTs, and indeed in most areas of physics. That is, we imagine that one has some individuating schema that allows for an identification of systems (equivalently, degrees of freedom) and an identification of the set of experimental procedures that pertain to these systems. In many physical contexts, especially those where experimentalists have a good deal of control, it is simple to identify systems and experimental procedures which act on them---there is little ambiguity in what is meant by the polarization degree of freedom of a photon, or by transformations on it. However, {\em formalizing} the schema by which physicists identify systems on the basis of operational statistics is a more subtle matter~\cite{Giuliosubsystems,kramer2018operational,mazurek2017experimentally,grabowecky2021experimentally}. 

When one goes beyond prepare-measure scenarios, stronger causal assumptions are generally required. As we saw in Sec.~\ref{sec:LabNotebook}, one may also make assumptions about the subsystem structure of composite systems in order to obtain the strongest possible constraints from noncontextuality. Given that subsystem structure can be understood as a type of causal structure (see Appendix B of Ref.~\cite{schmid2020unscrambling}), this is another type of causal assumption. 

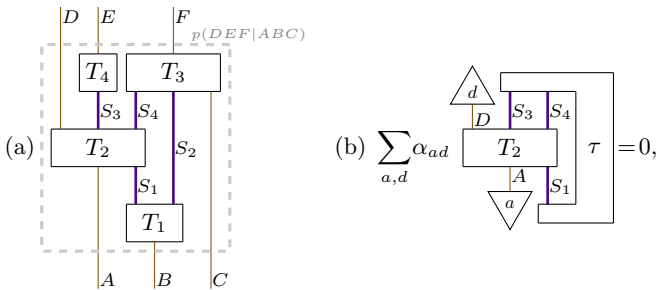
\begin{figure}[htbp!]
\begin{center}
(a) $%
\begin{tikzpicture}
	\begin{pgfonlayer}{nodelayer}
		\node [style=none] (6) at (-0.75, -1.5) {};
		\node [style=none] (7) at (-0.75, -2.5) {};
		\node [style=none] (8) at (0.75, -2.5) {};
		\node [style=none] (9) at (0.75, -1.5) {};
		\node [style=none] (10) at (0, -2.5) {};
		\node [style=none] (11) at (0, -3.75) {};
		\node [style=right label] (15) at (0, -3.5) {$B$};
		\node [style=none] (19) at (0, -2) {$T_1$};
		\node [style=none] (20) at (-0.5, -0.5) {};
		\node [style=none] (21) at (-0.5, -1.5) {};
		\node [style=right label] (22) at (-0.5, -1) {$S_1$};
		\node [style=none] (23) at (0.5, 1.5) {};
		\node [style=none] (24) at (0.5, -1.5) {};
		\node [style=right label] (25) at (0.5, 0) {$S_2$};
		\node [style=none] (26) at (-2.75, 0.5) {};
		\node [style=none] (27) at (-2.75, -0.5) {};
		\node [style=none] (28) at (-0.25, -0.5) {};
		\node [style=none] (29) at (-0.25, 0.5) {};
		\node [style=none] (30) at (-1.5, -0.5) {};
		\node [style=none] (31) at (-1.5, -3.75) {};
		\node [style=right label] (32) at (-1.5, -3.5) {$A$};
		\node [style=none] (33) at (-1.5, 0) {$T_2$};
		\node [style=none] (37) at (-0.5, 1.5) {};
		\node [style=none] (38) at (-0.5, 0.5) {};
		\node [style=right label] (39) at (-0.5, 1) {$S_4$};
		\node [style=none] (40) at (-1.5, 1.5) {};
		\node [style=none] (41) at (-1.5, 0.5) {};
		\node [style=right label] (42) at (-1.5, 1) {$S_3$};
		\node [style=none] (43) at (1.5, 1.5) {};
		\node [style=none] (44) at (1.5, -3.75) {};
		\node [style=right label] (45) at (1.5, -3.5) {$C$};
		\node [style=none] (46) at (-0.75, 1.5) {};
		\node [style=none] (47) at (-0.75, 2.5) {};
		\node [style=none] (48) at (1.75, 2.5) {};
		\node [style=none] (49) at (1.75, 1.5) {};
		\node [style=none] (50) at (0.5, 2.5) {};
		\node [style=none] (51) at (0.5, 3.75) {};
		\node [style=right label] (52) at (0.5, 3.5) {$F$};
		\node [style=none] (53) at (0.5, 2) {$T_3$};
		\node [style=none] (55) at (-2, 2.5) {};
		\node [style=none] (56) at (-2, 1.5) {};
		\node [style=none] (57) at (-1, 1.5) {};
		\node [style=none] (58) at (-1, 2.5) {};
		\node [style=none] (60) at (-1.5, 2) {$T_4$};
		\node [style=none] (63) at (-1.5, 3.75) {};
		\node [style=none] (64) at (-1.5, 2.5) {};
		\node [style=right label] (65) at (-1.5, 3.5) {$E$};
		\node [style=none] (66) at (-2.5, 3.75) {};
		\node [style=none] (67) at (-2.5, 0.5) {};
		\node [style=right label] (68) at (-2.5, 3.5) {$D$};
		\node [style=none] (69) at (-3, 2.75) {};
		\node [style=none] (70) at (2, 2.75) {};
		\node [style=none] (71) at (2, -2.75) {};
		\node [style=none] (72) at (-3, -2.75) {};
		\node [style=label] (73) at (2.5, 3.1) {\color{gray}\tiny $p(\! D\! E\! F\! |\! A\! B\! C\! )$};
	\end{pgfonlayer}
	\begin{pgfonlayer}{edgelayer}
		\draw (6.center) to (9.center);
		\draw (9.center) to (8.center);
		\draw (8.center) to (7.center);
		\draw (7.center) to (6.center);
		\draw [cWire] (11.center) to (10.center);
		\draw [qWire, in=-90, out=90] (21.center) to (20.center);
		\draw [qWire, in=-90, out=90] (24.center) to (23.center);
		\draw (26.center) to (29.center);
		\draw (29.center) to (28.center);
		\draw (28.center) to (27.center);
		\draw (27.center) to (26.center);
		\draw [cWire] (31.center) to (30.center);
		\draw [qWire, in=-90, out=90] (38.center) to (37.center);
		\draw [qWire, in=-90, out=90] (41.center) to (40.center);
		\draw [cWire] (44.center) to (43.center);
		\draw (46.center) to (49.center);
		\draw (49.center) to (48.center);
		\draw (48.center) to (47.center);
		\draw (47.center) to (46.center);
		\draw [cWire] (51.center) to (50.center);
		\draw (55.center) to (58.center);
		\draw (58.center) to (57.center);
		\draw (57.center) to (56.center);
		\draw (56.center) to (55.center);
		\draw [cWire, in=-90, out=90] (64.center) to (63.center);
		\draw [cWire, in=-90, out=90] (67.center) to (66.center);
		\draw [thick gray dashed edge] (70.center)
			 to (69.center)
			 to (72.center)
			 to (71.center)
			 to cycle;
	\end{pgfonlayer}
\end{tikzpicture}}$\hspace{2mm} (b) \ $\displaystyle{\sum_{a, d}} \alpha_{ad} %
\begin{tikzpicture}
	\begin{pgfonlayer}{nodelayer}
		\node [style=none] (20) at (-0.5, -0.5) {};
		\node [style=none] (21) at (-0.5, -1.5) {};
		\node [style=right label] (22) at (-0.5, -1) {$S_1$};
		\node [style=none] (26) at (-2.75, 0.5) {};
		\node [style=none] (27) at (-2.75, -0.5) {};
		\node [style=none] (28) at (-0.25, -0.5) {};
		\node [style=none] (29) at (-0.25, 0.5) {};
		\node [style=none] (30) at (-1.5, -0.5) {};
		\node [style=point] (31) at (-1.5, -1.5) {$a$};
		\node [style=right label] (32) at (-1.5, -0.75) {$A$};
		\node [style=none] (33) at (-1.5, 0) {$T_2$};
		\node [style=none] (37) at (-0.5, 1.5) {};
		\node [style=none] (38) at (-0.5, 0.5) {};
		\node [style=right label] (39) at (-0.5, 1) {$S_4$};
		\node [style=none] (40) at (-1.5, 1.5) {};
		\node [style=none] (41) at (-1.5, 0.5) {};
		\node [style=right label] (42) at (-1.5, 1) {$S_3$};
		\node [style=copoint] (66) at (-2.5, 1.5) {$d$};
		\node [style=none] (67) at (-2.5, 0.5) {};
		\node [style=right label] (68) at (-2.5, 0.75) {$D$};
		\node [style=none] (69) at (-1.75, 1.5) {};
		\node [style=none] (70) at (-1.75, 2) {};
		\node [style=none] (71) at (0.25, 1.5) {};
		\node [style=none] (72) at (0.25, -1.5) {};
		\node [style=none] (73) at (-0.75, -1.5) {};
		\node [style=none] (74) at (-0.75, -2) {};
		\node [style=none] (75) at (1.25, -2) {};
		\node [style=none] (76) at (1.25, 2) {};
		\node [style=none] (77) at (0.75, 0) {$\tau$};
	\end{pgfonlayer}
	\begin{pgfonlayer}{edgelayer}
		\draw [qWire, in=-90, out=90] (21.center) to (20.center);
		\draw (26.center) to (29.center);
		\draw (29.center) to (28.center);
		\draw (28.center) to (27.center);
		\draw (27.center) to (26.center);
		\draw [cWire] (31) to (30.center);
		\draw [qWire, in=-90, out=90] (38.center) to (37.center);
		\draw [qWire, in=-90, out=90] (41.center) to (40.center);
		\draw [cWire, in=-90, out=90] (67.center) to (66);
		\draw (69.center) to (71.center);
		\draw (71.center) to (72.center);
		\draw (72.center) to (73.center);
		\draw (73.center) to (74.center);
		\draw (74.center) to (75.center);
		\draw (75.center) to (76.center);
		\draw (76.center) to (70.center);
		\draw (70.center) to (69.center);
	\end{pgfonlayer}
\end{tikzpicture}} = 0,$ 
\caption{To determine if the statistics $p(DEF|ABC)$ generated by some GPT circuit are classically-explainable or not, one must look at the operational  identities holding among different processes that could appear in any given gate within the circuit. 
Thus, these operational identities can only be defined relative to the circuit structure. For instance, an operational identity  between the different possibilities for the gate taking $S_1$ to $S_3, S_4$ (indexed by setting $A$ and outcome $D$)  
 can be written as in (b).
 }
\label{circuit}
\end{center}
\end{figure}

Let us now set up a more general example, where one imagines an experiment, perhaps as a subroutine of a quantum computation.
 We will take Fig.~\ref{circuit}(a) as our working example.  The belief that the experiment is governed by this circuit is a {\em causal hypothesis}, and it contains a great deal more information than the operational statistics $p(DEF|ABC)$ on their own. The circuit diagram represents a commitment to the existence of a number of systems ($S_1, S_2, S_3, S_4$), represented as bold wires in the circuit, together with transformations ($T_1, T_2, T_3, T_4$), represented as gates in the circuit ,  and where the transformations may be chosen via classical setting variables $(A, B, C)$ and may output classical outcome variables $(D, E, F)$. 
The assumption that an experiment can be decomposed in this manner furthermore relies on the assumption that these transformations are autonomous in the sense that any one can be varied independently of the others. Note that an assumption of autonomy of causal mechanisms is also central to the framework for causal modelling used in the classical sphere~\cite{pearl2009causality,spirtes2000causation} and allows for inferences about counterfactual questions, such as how the observed outcome statistics would have been different if one had modified the circuit in a particular manner. From the raw operational statistics $p(DEF|ABC)$ on their own, there is generally no way 
to verify that a given causal hypothesis is correct. However, constructing causal hypotheses is one of the central tasks of science, and one which is increasingly studied in a formal manner, both in classical and quantum contexts~\cite{pearl2009causality,spirtes2000causation,costa2016quantum,allen2017quantum,Barrett2019}.

Consider now how noncontextuality arguments proceed under the assumption that Fig.~\ref{circuit}(a) describes the causal structure. We will do so from both of the two different perspectives on noncontextuality, described in Sec.~\ref{opequiv} and Sec.~\ref{gptnc} respectively.

Consider first the schema of Sec.~\ref{opequiv}, wherein one begins by identifying operational identities holding among the processes generating the observed correlations. This requires one to consider each possible circuit element individually. 
Obviously, this requires knowing the input and output systems of each gate
  in the circuit, which is information contained in the causal hypothesis. One can see this graphically using the notion of a {\em tester}, e.g., the comb $\tau$ in Fig.~\ref{circuit}(b); we refer the reader to Refs.~\cite{schmid2020structure,chiribella2010probabilistic} for details.

The second perspective is essentially an extension of the schema of Sec.~\ref{gptnc} from prepare-measure scenarios to arbitrary causal structures. It gives a holistic characterization of when an arbitrary GPT circuit admits of a classical explanation. As was shown in Ref.~\cite{schmid2020structure}, a GPT circuit admits of a classical explanation (in the sense that the operational theory which it describes admits of a noncontextual model) if and only if one can find a linear, diagram-preserving map taking it into the process theory of substochastic matrices, while preserving the predicted correlations. This is equivalent to asking if a positive quasiprobability representation exists for the GPT in question~\cite{schmid2020structure,negativity}.

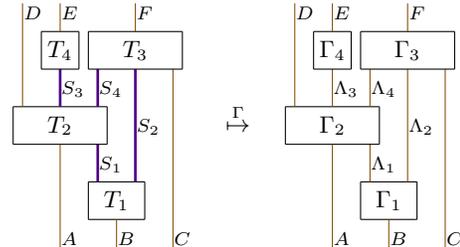
\begin{figure}[htbp!]
\begin{center}
$%
\begin{tikzpicture}
	\begin{pgfonlayer}{nodelayer}
		\node [style=none] (6) at (-0.75, -1.5) {};
		\node [style=none] (7) at (-0.75, -2.5) {};
		\node [style=none] (8) at (0.75, -2.5) {};
		\node [style=none] (9) at (0.75, -1.5) {};
		\node [style=none] (10) at (0, -2.5) {};
		\node [style=none] (11) at (0, -3.25) {};
		\node [style=right label] (15) at (0, -3) {$B$};
		\node [style=none] (19) at (0, -2) {$T_1$};
		\node [style=none] (20) at (-0.5, -0.5) {};
		\node [style=none] (21) at (-0.5, -1.5) {};
		\node [style=right label] (22) at (-0.5, -1) {$S_1$};
		\node [style=none] (23) at (0.5, 1.5) {};
		\node [style=none] (24) at (0.5, -1.5) {};
		\node [style=right label] (25) at (0.5, 0) {$S_2$};
		\node [style=none] (26) at (-2.75, 0.5) {};
		\node [style=none] (27) at (-2.75, -0.5) {};
		\node [style=none] (28) at (-0.25, -0.5) {};
		\node [style=none] (29) at (-0.25, 0.5) {};
		\node [style=none] (30) at (-1.5, -0.5) {};
		\node [style=none] (31) at (-1.5, -3.25) {};
		\node [style=right label] (32) at (-1.5, -3) {$A$};
		\node [style=none] (33) at (-1.5, 0) {$T_2$};
		\node [style=none] (37) at (-0.5, 1.5) {};
		\node [style=none] (38) at (-0.5, 0.5) {};
		\node [style=right label] (39) at (-0.5, 1) {$S_4$};
		\node [style=none] (40) at (-1.5, 1.5) {};
		\node [style=none] (41) at (-1.5, 0.5) {};
		\node [style=right label] (42) at (-1.5, 1) {$S_3$};
		\node [style=none] (43) at (1.5, 1.5) {};
		\node [style=none] (44) at (1.5, -3.25) {};
		\node [style=right label] (45) at (1.5, -3) {$C$};
		\node [style=none] (46) at (-0.75, 1.5) {};
		\node [style=none] (47) at (-0.75, 2.5) {};
		\node [style=none] (48) at (1.75, 2.5) {};
		\node [style=none] (49) at (1.75, 1.5) {};
		\node [style=none] (50) at (0.5, 2.5) {};
		\node [style=none] (51) at (0.5, 3.25) {};
		\node [style=right label] (52) at (0.5, 3) {$F$};
		\node [style=none] (53) at (0.5, 2) {$T_3$};
		\node [style=none] (55) at (-2, 2.5) {};
		\node [style=none] (56) at (-2, 1.5) {};
		\node [style=none] (57) at (-1, 1.5) {};
		\node [style=none] (58) at (-1, 2.5) {};
		\node [style=none] (60) at (-1.5, 2) {$T_4$};
		\node [style=none] (63) at (-1.5, 3.25) {};
		\node [style=none] (64) at (-1.5, 2.5) {};
		\node [style=right label] (65) at (-1.5, 3) {$E$};
		\node [style=none] (66) at (-2.5, 3.25) {};
		\node [style=none] (67) at (-2.5, 0.5) {};
		\node [style=right label] (68) at (-2.5, 3) {$D$};
	\end{pgfonlayer}
	\begin{pgfonlayer}{edgelayer}
		\draw (6.center) to (9.center);
		\draw (9.center) to (8.center);
		\draw (8.center) to (7.center);
		\draw (7.center) to (6.center);
		\draw [cWire] (11.center) to (10.center);
		\draw [qWire, in=-90, out=90] (21.center) to (20.center);
		\draw [qWire, in=-90, out=90] (24.center) to (23.center);
		\draw (26.center) to (29.center);
		\draw (29.center) to (28.center);
		\draw (28.center) to (27.center);
		\draw (27.center) to (26.center);
		\draw [cWire] (31.center) to (30.center);
		\draw [qWire, in=-90, out=90] (38.center) to (37.center);
		\draw [qWire, in=-90, out=90] (41.center) to (40.center);
		\draw [cWire] (44.center) to (43.center);
		\draw (46.center) to (49.center);
		\draw (49.center) to (48.center);
		\draw (48.center) to (47.center);
		\draw (47.center) to (46.center);
		\draw [cWire] (51.center) to (50.center);
		\draw (55.center) to (58.center);
		\draw (58.center) to (57.center);
		\draw (57.center) to (56.center);
		\draw (56.center) to (55.center);
		\draw [cWire, in=-90, out=90] (64.center) to (63.center);
		\draw [cWire, in=-90, out=90] (67.center) to (66.center);
	\end{pgfonlayer}
\end{tikzpicture}} \quad \stackrel{\Gamma}{\mapsto}\quad 
\ %
\begin{tikzpicture}
	\begin{pgfonlayer}{nodelayer}
		\node [style=none] (6) at (-0.75, -1.5) {};
		\node [style=none] (7) at (-0.75, -2.5) {};
		\node [style=none] (8) at (0.75, -2.5) {};
		\node [style=none] (9) at (0.75, -1.5) {};
		\node [style=none] (10) at (0, -2.5) {};
		\node [style=none] (11) at (0, -3.25) {};
		\node [style=right label] (15) at (0, -3) {$B$};
		\node [style=none] (19) at (0, -2) {$\Gamma_1$};
		\node [style=none] (20) at (-0.5, -0.5) {};
		\node [style=none] (21) at (-0.5, -1.5) {};
		\node [style=right label] (22) at (-0.5, -1) {$\Lambda_1$};
		\node [style=none] (23) at (0.5, 1.5) {};
		\node [style=none] (24) at (0.5, -1.5) {};
		\node [style=right label] (25) at (0.5, 0) {$\Lambda_2$};
		\node [style=none] (26) at (-2.75, 0.5) {};
		\node [style=none] (27) at (-2.75, -0.5) {};
		\node [style=none] (28) at (-0.25, -0.5) {};
		\node [style=none] (29) at (-0.25, 0.5) {};
		\node [style=none] (30) at (-1.5, -0.5) {};
		\node [style=none] (31) at (-1.5, -3.25) {};
		\node [style=right label] (32) at (-1.5, -3) {$A$};
		\node [style=none] (33) at (-1.5, 0) {$\Gamma_2$};
		\node [style=none] (37) at (-0.5, 1.5) {};
		\node [style=none] (38) at (-0.5, 0.5) {};
		\node [style=right label] (39) at (-0.5, 1) {$\Lambda_4$};
		\node [style=none] (40) at (-1.5, 1.5) {};
		\node [style=none] (41) at (-1.5, 0.5) {};
		\node [style=right label] (42) at (-1.5, 1) {$\Lambda_3$};
		\node [style=none] (43) at (1.5, 1.5) {};
		\node [style=none] (44) at (1.5, -3.25) {};
		\node [style=right label] (45) at (1.5, -3) {$C$};
		\node [style=none] (46) at (-0.75, 1.5) {};
		\node [style=none] (47) at (-0.75, 2.5) {};
		\node [style=none] (48) at (1.75, 2.5) {};
		\node [style=none] (49) at (1.75, 1.5) {};
		\node [style=none] (50) at (0.5, 2.5) {};
		\node [style=none] (51) at (0.5, 3.25) {};
		\node [style=right label] (52) at (0.5, 3) {$F$};
		\node [style=none] (53) at (0.5, 2) {$\Gamma_3$};
		\node [style=none] (55) at (-2, 2.5) {};
		\node [style=none] (56) at (-2, 1.5) {};
		\node [style=none] (57) at (-1, 1.5) {};
		\node [style=none] (58) at (-1, 2.5) {};
		\node [style=none] (60) at (-1.5, 2) {$\Gamma_4$};
		\node [style=none] (63) at (-1.5, 3.25) {};
		\node [style=none] (64) at (-1.5, 2.5) {};
		\node [style=right label] (65) at (-1.5, 3) {$E$};
		\node [style=none] (66) at (-2.5, 3.25) {};
		\node [style=none] (67) at (-2.5, 0.5) {};
		\node [style=right label] (68) at (-2.5, 3) {$D$};
	\end{pgfonlayer}
	\begin{pgfonlayer}{edgelayer}
		\draw (6.center) to (9.center);
		\draw (9.center) to (8.center);
		\draw (8.center) to (7.center);
		\draw (7.center) to (6.center);
		\draw [cWire] (11.center) to (10.center);
		\draw [cWire, in=-90, out=90] (21.center) to (20.center);
		\draw [cWire, in=-90, out=90] (24.center) to (23.center);
		\draw (26.center) to (29.center);
		\draw (29.center) to (28.center);
		\draw (28.center) to (27.center);
		\draw (27.center) to (26.center);
		\draw [cWire] (31.center) to (30.center);
		\draw [cWire, in=-90, out=90] (38.center) to (37.center);
		\draw [cWire, in=-90, out=90] (41.center) to (40.center);
		\draw [cWire] (44.center) to (43.center);
		\draw (46.center) to (49.center);
		\draw (49.center) to (48.center);
		\draw (48.center) to (47.center);
		\draw (47.center) to (46.center);
		\draw [cWire] (51.center) to (50.center);
		\draw (55.center) to (58.center);
		\draw (58.center) to (57.center);
		\draw (57.center) to (56.center);
		\draw (56.center) to (55.center);
		\draw [cWire, in=-90, out=90] (64.center) to (63.center);
		\draw [cWire, in=-90, out=90] (67.center) to (66.center);
	\end{pgfonlayer}
\end{tikzpicture}}$ 
\caption{As shown in Ref.~\cite{schmid2020structure}, the statistics $p(DEF|ABC)$ generated by some GPT circuit are explainable within a noncontextual ontological model $\Gamma$ if and only if there exists a linear map from the GPT circuit into a circuit of the same form but where all systems are classical random variables, and where all transformations are substochastic maps.}
\label{maptoss}
\end{center}
\end{figure}

This is illustrated schematically in Fig.~\ref{maptoss}. Clearly, one can only evaluate nonclassicality in this manner if one already has a circuit---a causal hypothesis---in mind. 

Given a particular causal hypothesis, one always has the option to lump together processes to obtain a coarse-grained description; for example, in a prepare-transform-measure scenario on a single system, one may lump together the transformation and the measurement to reduce the scenario to an effective prepare-measure scenario. 
While this lumping of circuit elements can sometimes simplify one's analysis, it can also prevent one from deducing the {\em complete} consequences of noncontextuality relative to the causal hypothesis. 

Take for example the stabilizer theory of a single qubit. In this operational theory,  one can find proofs of contextual prepare-transform-measure scenarios, but one {\em cannot} find any such proofs in prepare-measure scenarios~\cite{Lillystone2019}. And yet, the composition of any transformation with any state (or measurement) in the theory yields another state (or measurement) in the theory. Thus, it follows that when one 
lumps together the transformation with either the state or the effect, the effective scenario that results is one which  admits 
 of a noncontextual model. This is simply a consequence of the choice to only carry out a coarse-grained analysis. In general, one must consider the ontological representation of each circuit element individually in order to determine necessary and sufficient conditions for noncontextuality, relative to the (fine-grained) causal hypothesis.

The considerations of this section raise the question of how one decides between different causal hypotheses, a question to which we turn in the next section.

\subsection{When one should assume diagram preservation with respect to the standard quantum circuit}\label{whichcircuit}

As mentioned earlier, constructing a causal hypothesis for some observed phenomena is a difficult but central scientific task. We will now argue that---at least in various branches of experimental quantum information processing, where one has a good deal of control over the systems in question---it is typically straightforward to write down {\em the} quantum circuit associated with a given experiment or protocol. We argue that the structure of this circuit provides the natural causal hypothesis, and if one seeks realist explanations of the observed data, one should demand that the explanation respects this causal hypothesis.

On what grounds do physicists ever associate a system with some operational procedures carried out in a laboratory? We poke and we prod at the world until we identify meaningful loci of intervention. With enough experimentation, we eventually distill out meaningful notions of systems (like electrons, photons, etc), and we imagine that these systems have properties which are prepared, measured, and transformed by our interventions. These systems, then, are the most natural candidates for causal mediaries, and the standard quantum circuit describing the experiment is the natural candidate for the causal structure.
 In other words, this causal hypothesis is the most natural culmination of all the evidence gathered to date for how one can break up an experiment into localized systems and autonomous transformations on them. To postulate any other causal structure is a more radical move, and requires special justification.

Consider for example the standard Bell scenario. The standard quantum circuit for this scenario invokes a common cause which sets up correlations between the local measurements performed by the two parties. The conservative causal hypothesis is that the causal structure has the same form at the ontological level, and it is precisely this assumption that leads to Bell inequalities. 

The generalization of this line of reasoning to arbitrary causal scenarios is given by the rapidly growing field of causal compatibility inequalities~\cite{wood2015lesson,FritzBeyondBellI,tavakoli2021bell,ChavesNetworks2021}, where one demonstrates nonclassicality by showing that quantum circuits of a given causal structure are capable of generating a broader set of correlations than classical ones. Such arguments {\em also} rely on the assumption that an experiment associated with a given quantum circuit is in fact a faithful realization of the causal structure described by the quantum circuit.

Of course, some physicists (such as proponents of Bohmian mechanics) do believe in superluminal causal influences, and so would advocate for a causal hypothesis which does {\em not} mirror the structure of the standard quantum circuit. So this is not to say that there is {\em no} sense in considering nonstandard causal hypotheses; however, interpretations which make radical causal assumptions are made less compelling as a consequence. Moreover, these nonstandard causal hypotheses are typically endorsed by those who do not believe that there is any other way out of no-go theorems like Bell's, but, as we argue in the next section, we believe that there {\em are}, in fact, other  ways out.

Another reason to demand that the ontological representation of an experiment respect the conservative causal structure is that representations using radical causal structures typically {\em overfit} the data. For example, explanations of Bell inequality violations which appeal to superluminal causation are often general enough to allow for signalling correlations, and overfitting is generally a consequence of this~\cite{daley2021experimentally}. Alternatively, one can avoid this overfitting by imposing restrictions on the scope of causal mechanisms allowed in the model, but these restrictions generally lead to violations of noncontextuality~\cite{fromBelltoNC}. 

A motivation for studying ontological representations is to the search for deeper explanations for our experiments and theories. The most natural explanations are those with the most conservative assumptions about the causal structure, which we have argued correspond to the standard quantum circuit representation. It follows that one should focus one's attention on ontological models that respect the structure of the quantum circuit. This desiderata is captured by demanding diagram-preservation relative to the standard quantum circuit. 

\subsection{What to do in light of the failure of noncontextuality in quantum theory}

For experiments described by operational quantum theory, one cannot necessarily find ontological representations that respect the conservative causal structure and are noncontextual.  There are by now many proofs of this fact, spanning a variety of physical scenarios~\cite{PP1,PP2,POM,parable,AWV,robust,operationalks,RAC,schmid2018contextual,KLP19,saha2019preparation,Lillystone2019,RAC2,contextmetrology,cloningcontext,Yadavalli2020,selby2023incompatibility,selby2023accessible,Roch2021,Flatt2021,Schmid2022Stabilizer,Schmid2024reviewreformulation}. What should one conclude when faced with these no-go theorems?

There are three natural possibilities.
The first is to imagine that quantum theory is not the true theory of nature and that the operational equivalences that have been observed to date are not operational equivalences in the true theory, which might then still be consistent with noncontextuality.
The second is to simply bite the bullet and grant that nature is described by a contextual ontological model. This is the route, for instance, that advocates of Bohmian mechanics endorse.   The third is to relax some of the background assumptions going into proofs of noncontextuality, such as the assumptions built into the framework of ontological models, in a manner that allows one 
to maintain the spirit of noncontextuality.

The first response above is, in our view, unlikely to be the correct resolution to the problem.  For one,  it can be shown that the operational equivalences arising in quantum theory will {\em continue to hold} in a post-quantum theory, if one grants a few reasonable physical assumptions regarding the sense in which quantum theory emerges from this postquantum theory via a decoherence-like process~\cite{muller2023testing,NCsubsystems}. Also, in the case where the operational equivalences follow from the lack of signalling between space-like separated regions, such as in a Bell experiment, this sort of response requires one to imagine that the true post-quantum theory is one that allows for superluminal signalling and hence conflicts with relativity theory, a possibility that we take to be unlikely. 

We find the second response to be unsatisfactory because contextual theories lack much of the explanatory power that are provided by noncontextual theories. For instance, if one considers the operational phenomenology of the odd-dimensional stabilizer subtheory of quantum theory, then the Bohmian account of this phenomenology is far more convoluted and counterintuitive than the description provided by the Spekkens toy theory~\cite{spekkens2007evidence}, or equivalently, Gross's discrete Wigner representation~\cite{gross2006hudson}.

This leaves the third possible response, that one must consider modifying the framework of ontological models in such a way that one can construct a realist description of quantum theory that salvages the spirit of noncontextuality. This is easier said than done, since the standard framework of ontological models is an extremely general and compelling framework for providing realist explanations, and it is unclear how to modify it while retaining these features. 
 Nonetheless, we believe that this is the correct response, and 
 first steps in this direction can be found in Ref.~\cite{schmid2020unscrambling} (see also Ref.~\cite{Pirsa_omelette}).
 In particular,  for the special class of contextuality experiments that are Bell experiments, while a standard response to Bell-inequality violations is to concede that nature allows superluminal causes (relativity be damned), the third type of response asks one to instead question the background assumptions going into Bell-like no-go theorems. 
If one modifies the framework of causal modelling 
that underlies these no-go theorems, then one can hope to find an intrinsically quantum notion of causation~\cite{costa2016quantum,allen2017quantum,Barrett2019} that reproduces 
 the quantum predictions in Bell scenarios while preserving the spirit of locality. More specifically, the aim of such works is to explain Bell violations as consequences of nonclassical common causes~\cite{cavalcanti2014modifications,allen2017quantum,Wolfe2020quantifyingbell,Schmid2020typeindependent,Schmid2023understanding} rather than superluminal causes. 

Thus, our preferred response to both noncontextuality no-go theorems and Bell-like 
no-go theorems is to
 devise a more general notion of {\em nonclassical realism} that allows us to give causal explanations of observed correlations in a manner that is consistent with the Leibnizian methodological principle~\cite{schmid2020unscrambling,Pirsa_omelette}.

\tocless\section{Acknowledgements}

We thank Elie Wolfe, Ana Bel\'en Sainz, and Ravi Kunjwal for useful discussions, especially regarding operational equivalences among composite systems and regarding the PR box example. 
We also thank Matt Pusey for useful discussions, and thank L\'idia del Rio for feedback on Section III, and indeed for motivating us to write it (whether or not that was intentional!). JHS was supported by the National Science Centre, Poland (Opus project, Categorical Foundations of the Non-Classicality of Nature, project no. 2021/41/B/ST2/03149). DS was supported by the Foundation for Polish Science (IRAP project, ICTQT, contract no. MAB/2018/5, co-financed by EU within Smart Growth Operational Programme). DS was also supported by the National Science Centre, Poland (Opus project,
Categorical Foundations of the Non-Classicality of Nature,
project no. 2021/41/B/ST2/03149).

All diagrams were prepared using TikZit.

\bibliographystyle{apsrev4-2-wolfe}
\setlength{\bibsep}{3pt plus 3pt minus 2pt}
\bibliography{bib.bib}

\begin{thebibliography}{83}%
\makeatletter
\providecommand \@ifxundefined [1]{%
 \@ifx{#1\undefined}
}%
\providecommand \@ifnum [1]{%
 \ifnum #1\expandafter \@firstoftwo
 \else \expandafter \@secondoftwo
 \fi
}%
\providecommand \@ifx [1]{%
 \ifx #1\expandafter \@firstoftwo
 \else \expandafter \@secondoftwo
 \fi
}%
\providecommand \natexlab [1]{#1}%
\providecommand \enquote  [1]{``#1''}%
\providecommand \bibnamefont  [1]{#1}%
\providecommand \bibfnamefont [1]{#1}%
\providecommand \citenamefont [1]{#1}%
\providecommand \href@noop [0]{\@secondoftwo}%
\providecommand \href [0]{\begingroup \@sanitize@url \@href}%
\providecommand \@href[1]{\@@startlink{#1}\@@href}%
\providecommand \@@href[1]{\endgroup#1\@@endlink}%
\providecommand \@sanitize@url [0]{\catcode `\\12\catcode `\$12\catcode
  `\&12\catcode `\#12\catcode `\^12\catcode `\_12\catcode `\%12\relax}%
\providecommand \@@startlink[1]{}%
\providecommand \@@endlink[0]{}%
\providecommand \url  [0]{\begingroup\@sanitize@url \@url }%
\providecommand \@url [1]{\endgroup\@href {#1}{\urlprefix }}%
\providecommand \urlprefix  [0]{URL }%
\providecommand \Eprint [0]{\href }%
\providecommand \doibase [0]{https://doi.org/}%
\providecommand \selectlanguage [0]{\@gobble}%
\providecommand \bibinfo  [0]{\@secondoftwo}%
\providecommand \bibfield  [0]{\@secondoftwo}%
\providecommand \translation [1]{[#1]}%
\providecommand \BibitemOpen [0]{}%
\providecommand \bibitemStop [0]{}%
\providecommand \bibitemNoStop [0]{.\EOS\space}%
\providecommand \EOS [0]{\spacefactor3000\relax}%
\providecommand \BibitemShut  [1]{\csname bibitem#1\endcsname}%
\let\auto@bib@innerbib\@empty
\bibitem [{\citenamefont {Spekkens}(2005)}]{gencontext}%
  \BibitemOpen
  \bibfield  {author} {\bibinfo {author} {\bibfnamefont {R.~W.}\ \bibnamefont
  {Spekkens}},\ }\href {http://link.aps.org/doi/10.1103/PhysRevA.71.052108}
  {\bibfield  {journal} {\bibinfo  {journal} {Phys. Rev. A}\ }\textbf {\bibinfo
  {volume} {71}},\ \bibinfo {pages} {052108} (\bibinfo {year}
  {2005})}\BibitemShut {NoStop}%
\bibitem [{\citenamefont {Kochen}\ and\ \citenamefont {Specker}(1967)}]{KS}%
  \BibitemOpen
  \bibfield  {author} {\bibinfo {author} {\bibfnamefont {S.}~\bibnamefont
  {Kochen}}\ and\ \bibinfo {author} {\bibfnamefont {E.}~\bibnamefont
  {Specker}},\ }\href {https://doi.org/10.1007/978-94-010-1795-4\_17}
  {\bibfield  {journal} {\bibinfo  {journal} {J. Math. \& Mech.}\ }\textbf
  {\bibinfo {volume} {17}},\ \bibinfo {pages} {59} (\bibinfo {year} {1967})},\
  \bibinfo {note} {also available from the
  \href[pdfnewwindow]{http://www.iumj.indiana.edu/IUMJ/fulltext.php?year=1968\&volume=17\&artid=17004}{Indiana
  Univ. Math. J.}}\BibitemShut {Stop}%
\bibitem [{\citenamefont {{Spekkens}}(2019)}]{Leibniz}%
  \BibitemOpen
  \bibfield  {author} {\bibinfo {author} {\bibfnamefont {R.~W.}\ \bibnamefont
  {{Spekkens}}},\ }\href@noop {} {\enquote {\bibinfo {title} {{The ontological
  identity of empirical indiscernibles: Leibniz's methodological principle and
  its significance in the work of Einstein}},}\ } (\bibinfo {year} {2019}),\
  \Eprint {https://arxiv.org/abs/1909.04628} {arXiv:1909.04628
  [physics.hist-ph]} \BibitemShut {NoStop}%
\bibitem [{\citenamefont {Schmid}(2021)}]{schmid2021guiding}%
  \BibitemOpen
  \bibfield  {author} {\bibinfo {author} {\bibfnamefont {D.}~\bibnamefont
  {Schmid}},\ }\href {http://hdl.handle.net/10012/17136} {\bibfield  {journal}
  {\bibinfo  {journal} {PhD thesis, University of Waterloo}\ } (\bibinfo {year}
  {2021})}\BibitemShut {NoStop}%
\bibitem [{\citenamefont {Selby}\ \emph {et~al.}(2024)\citenamefont {Selby},
  \citenamefont {Wolfe}, \citenamefont {Schmid}, \citenamefont {Sainz},\ and\
  \citenamefont {Rossi}}]{selby2024linear}%
  \BibitemOpen
  \bibfield  {author} {\bibinfo {author} {\bibfnamefont {J.~H.}\ \bibnamefont
  {Selby}}, \bibinfo {author} {\bibfnamefont {E.}~\bibnamefont {Wolfe}},
  \bibinfo {author} {\bibfnamefont {D.}~\bibnamefont {Schmid}}, \bibinfo
  {author} {\bibfnamefont {A.~B.}\ \bibnamefont {Sainz}},\ and\ \bibinfo
  {author} {\bibfnamefont {V.~P.}\ \bibnamefont {Rossi}},\ }\href
  {https://doi.org/10.1103/PhysRevLett.132.050202} {\bibfield  {journal}
  {\bibinfo  {journal} {Phys. Rev. Lett.}\ }\textbf {\bibinfo {volume} {132}},\
  \bibinfo {pages} {050202} (\bibinfo {year} {2024})}\BibitemShut {NoStop}%
\bibitem [{\citenamefont {Schmid}\ \emph
  {et~al.}(2020{\natexlab{a}})\citenamefont {Schmid}, \citenamefont {Selby},
  \citenamefont {Pusey},\ and\ \citenamefont {Spekkens}}]{schmid2020structure}%
  \BibitemOpen
  \bibfield  {author} {\bibinfo {author} {\bibfnamefont {D.}~\bibnamefont
  {Schmid}}, \bibinfo {author} {\bibfnamefont {J.~H.}\ \bibnamefont {Selby}},
  \bibinfo {author} {\bibfnamefont {M.~F.}\ \bibnamefont {Pusey}},\ and\
  \bibinfo {author} {\bibfnamefont {R.~W.}\ \bibnamefont {Spekkens}},\
  }\href@noop {} {\enquote {\bibinfo {title} {A structure theorem for
  generalized-noncontextual ontological models},}\ } (\bibinfo {year}
  {2020}{\natexlab{a}}),\ \Eprint {https://arxiv.org/abs/2005.07161}
  {arXiv:2005.07161 [quant-ph]} \BibitemShut {NoStop}%
\bibitem [{\citenamefont {Schmid}(2022{\natexlab{a}})}]{NCvid1}%
  \BibitemOpen
  \bibfield  {author} {\bibinfo {author} {\bibfnamefont {D.}~\bibnamefont
  {Schmid}},\ }\href {https://www.youtube.com/watch?v=M3qn3EHWdOg} {\enquote
  {\bibinfo {title} {Noncontextuality, part 1 ({Y}outube)},}\ } (\bibinfo
  {year} {2022}{\natexlab{a}})\BibitemShut {NoStop}%
\bibitem [{\citenamefont {Schmid}(2022{\natexlab{b}})}]{NCvid2}%
  \BibitemOpen
  \bibfield  {author} {\bibinfo {author} {\bibfnamefont {D.}~\bibnamefont
  {Schmid}},\ }\href {https://www.youtube.com/watch?v=htZPbBp8JBY} {\enquote
  {\bibinfo {title} {Noncontextuality, part 2 ({Y}outube)},}\ } (\bibinfo
  {year} {2022}{\natexlab{b}})\BibitemShut {NoStop}%
\bibitem [{\citenamefont {Schmid}(2022{\natexlab{c}})}]{NCvid3}%
  \BibitemOpen
  \bibfield  {author} {\bibinfo {author} {\bibfnamefont {D.}~\bibnamefont
  {Schmid}},\ }\href {https://www.youtube.com/watch?v=uyI3nPCeVNg} {\enquote
  {\bibinfo {title} {Noncontextuality, part 3 ({Y}outube)},}\ } (\bibinfo
  {year} {2022}{\natexlab{c}})\BibitemShut {NoStop}%
\bibitem [{\citenamefont {{Hardy}}(2001)}]{Hardy}%
  \BibitemOpen
  \bibfield  {author} {\bibinfo {author} {\bibfnamefont {L.}~\bibnamefont
  {{Hardy}}},\ }\href {https://arxiv.org/abs/quant-ph/0101012} {\bibfield
  {journal} {\bibinfo  {journal} {arXiv:quant-ph/0101012}\ } (\bibinfo {year}
  {2001})}\BibitemShut {NoStop}%
\bibitem [{\citenamefont {Barrett}(2007)}]{GPT_Barrett}%
  \BibitemOpen
  \bibfield  {author} {\bibinfo {author} {\bibfnamefont {J.}~\bibnamefont
  {Barrett}},\ }\href {https://doi.org/10.1103/PhysRevA.75.032304} {\bibfield
  {journal} {\bibinfo  {journal} {Phys. Rev. A}\ }\textbf {\bibinfo {volume}
  {75}},\ \bibinfo {pages} {032304} (\bibinfo {year} {2007})}\BibitemShut
  {NoStop}%
\bibitem [{\citenamefont {Chiribella}\ \emph {et~al.}(2010)\citenamefont
  {Chiribella}, \citenamefont {D'Ariano},\ and\ \citenamefont
  {Perinotti}}]{chiribella2010probabilistic}%
  \BibitemOpen
  \bibfield  {author} {\bibinfo {author} {\bibfnamefont {G.}~\bibnamefont
  {Chiribella}}, \bibinfo {author} {\bibfnamefont {G.~M.}\ \bibnamefont
  {D'Ariano}},\ and\ \bibinfo {author} {\bibfnamefont {P.}~\bibnamefont
  {Perinotti}},\ }\href {https://doi.org/10.1103/PhysRevA.81.062348} {\bibfield
   {journal} {\bibinfo  {journal} {Phys. Rev. A}\ }\textbf {\bibinfo {volume}
  {81}},\ \bibinfo {pages} {062348} (\bibinfo {year} {2010})}\BibitemShut
  {NoStop}%
\bibitem [{\citenamefont {Schmid}\ \emph
  {et~al.}(2021{\natexlab{a}})\citenamefont {Schmid}, \citenamefont {Selby},
  \citenamefont {Wolfe}, \citenamefont {Kunjwal},\ and\ \citenamefont
  {Spekkens}}]{SchmidGPT}%
  \BibitemOpen
  \bibfield  {author} {\bibinfo {author} {\bibfnamefont {D.}~\bibnamefont
  {Schmid}}, \bibinfo {author} {\bibfnamefont {J.~H.}\ \bibnamefont {Selby}},
  \bibinfo {author} {\bibfnamefont {E.}~\bibnamefont {Wolfe}}, \bibinfo
  {author} {\bibfnamefont {R.}~\bibnamefont {Kunjwal}},\ and\ \bibinfo {author}
  {\bibfnamefont {R.~W.}\ \bibnamefont {Spekkens}},\ }\href
  {https://doi.org/10.1103/PRXQuantum.2.010331} {\bibfield  {journal} {\bibinfo
   {journal} {PRX Quantum}\ }\textbf {\bibinfo {volume} {2}},\ \bibinfo {pages}
  {010331} (\bibinfo {year} {2021}{\natexlab{a}})}\BibitemShut {NoStop}%
\bibitem [{\citenamefont {Selby}\ \emph
  {et~al.}(2023{\natexlab{a}})\citenamefont {Selby}, \citenamefont {Schmid},
  \citenamefont {Wolfe}, \citenamefont {Sainz}, \citenamefont {Kunjwal},\ and\
  \citenamefont {Spekkens}}]{selby2023accessible}%
  \BibitemOpen
  \bibfield  {author} {\bibinfo {author} {\bibfnamefont {J.~H.}\ \bibnamefont
  {Selby}}, \bibinfo {author} {\bibfnamefont {D.}~\bibnamefont {Schmid}},
  \bibinfo {author} {\bibfnamefont {E.}~\bibnamefont {Wolfe}}, \bibinfo
  {author} {\bibfnamefont {A.~B.}\ \bibnamefont {Sainz}}, \bibinfo {author}
  {\bibfnamefont {R.}~\bibnamefont {Kunjwal}},\ and\ \bibinfo {author}
  {\bibfnamefont {R.~W.}\ \bibnamefont {Spekkens}},\ }\href
  {https://doi.org/10.1103/PhysRevA.107.062203} {\bibfield  {journal} {\bibinfo
   {journal} {Phys. Rev. A}\ }\textbf {\bibinfo {volume} {107}},\ \bibinfo
  {pages} {062203} (\bibinfo {year} {2023}{\natexlab{a}})}\BibitemShut
  {NoStop}%
\bibitem [{\citenamefont {Schmid}\ \emph
  {et~al.}(2021{\natexlab{b}})\citenamefont {Schmid}, \citenamefont {Selby},\
  and\ \citenamefont {Spekkens}}]{schmid2020unscrambling}%
  \BibitemOpen
  \bibfield  {author} {\bibinfo {author} {\bibfnamefont {D.}~\bibnamefont
  {Schmid}}, \bibinfo {author} {\bibfnamefont {J.~H.}\ \bibnamefont {Selby}},\
  and\ \bibinfo {author} {\bibfnamefont {R.~W.}\ \bibnamefont {Spekkens}},\
  }\href@noop {} {\enquote {\bibinfo {title} {Unscrambling the omelette of
  causation and inference: The framework of causal-inferential theories},}\ }
  (\bibinfo {year} {2021}{\natexlab{b}}),\ \Eprint
  {https://arxiv.org/abs/2009.03297} {arXiv:2009.03297 [quant-ph]} \BibitemShut
  {NoStop}%
\bibitem [{\citenamefont {Spekkens}(2014)}]{determinism}%
  \BibitemOpen
  \bibfield  {author} {\bibinfo {author} {\bibfnamefont {R.~W.}\ \bibnamefont
  {Spekkens}},\ }\href {https://doi.org/10.1007/s10701-014-9833-x} {\bibfield
  {journal} {\bibinfo  {journal} {Foundations of Physics}\ }\textbf {\bibinfo
  {volume} {44}},\ \bibinfo {pages} {1125} (\bibinfo {year}
  {2014})}\BibitemShut {NoStop}%
\bibitem [{\citenamefont {Mazurek}\ \emph {et~al.}(2017)\citenamefont
  {Mazurek}, \citenamefont {Pusey}, \citenamefont {Resch},\ and\ \citenamefont
  {Spekkens}}]{mazurek2017experimentally}%
  \BibitemOpen
  \bibfield  {author} {\bibinfo {author} {\bibfnamefont {M.~D.}\ \bibnamefont
  {Mazurek}}, \bibinfo {author} {\bibfnamefont {M.~F.}\ \bibnamefont {Pusey}},
  \bibinfo {author} {\bibfnamefont {K.~J.}\ \bibnamefont {Resch}},\ and\
  \bibinfo {author} {\bibfnamefont {R.~W.}\ \bibnamefont {Spekkens}},\ }\href
  {https://arxiv.org/abs/1710.05948} {\bibfield  {journal} {\bibinfo  {journal}
  {arXiv:1710.05948}\ } (\bibinfo {year} {2017})}\BibitemShut {NoStop}%
\bibitem [{\citenamefont {Grabowecky}\ \emph {et~al.}(2021)\citenamefont
  {Grabowecky}, \citenamefont {Pollack}, \citenamefont {Cameron}, \citenamefont
  {Spekkens},\ and\ \citenamefont {Resch}}]{grabowecky2021experimentally}%
  \BibitemOpen
  \bibfield  {author} {\bibinfo {author} {\bibfnamefont {M.}~\bibnamefont
  {Grabowecky}}, \bibinfo {author} {\bibfnamefont {C.}~\bibnamefont {Pollack}},
  \bibinfo {author} {\bibfnamefont {A.}~\bibnamefont {Cameron}}, \bibinfo
  {author} {\bibfnamefont {R.}~\bibnamefont {Spekkens}},\ and\ \bibinfo
  {author} {\bibfnamefont {K.}~\bibnamefont {Resch}},\ }\href@noop {} {\enquote
  {\bibinfo {title} {Experimentally bounding deviations from quantum theory for
  a photonic three-level system using theory-agnostic tomography},}\ }
  (\bibinfo {year} {2021}),\ \Eprint {https://arxiv.org/abs/2108.05864}
  {arXiv:2108.05864 [quant-ph]} \BibitemShut {NoStop}%
\bibitem [{\citenamefont {Schmid}(2019)}]{Pirsa_omelette}%
  \BibitemOpen
  \bibfield  {author} {\bibinfo {author} {\bibfnamefont {D.}~\bibnamefont
  {Schmid}},\ }\bibfield  {journal} {\bibinfo  {journal} {{PIRSA:19120030}}\
  }\href {https://doi.org/10.48660/19120030} {10.48660/19120030} (\bibinfo
  {year} {2019})\BibitemShut {NoStop}%
\bibitem [{\citenamefont {Popescu}\ and\ \citenamefont
  {Rohrlich}(1994)}]{Popescu1994}%
  \BibitemOpen
  \bibfield  {author} {\bibinfo {author} {\bibfnamefont {S.}~\bibnamefont
  {Popescu}}\ and\ \bibinfo {author} {\bibfnamefont {D.}~\bibnamefont
  {Rohrlich}},\ }\href {https://doi.org/10.1007/BF02058098} {\bibfield
  {journal} {\bibinfo  {journal} {Foundations of Physics}\ }\textbf {\bibinfo
  {volume} {24}},\ \bibinfo {pages} {379} (\bibinfo {year} {1994})}\BibitemShut
  {NoStop}%
\bibitem [{\citenamefont {Cavalcanti}\ \emph {et~al.}(2022)\citenamefont
  {Cavalcanti}, \citenamefont {Selby}, \citenamefont {Sikora}, \citenamefont
  {Galley},\ and\ \citenamefont {Sainz}}]{cavalcanti2022post}%
  \BibitemOpen
  \bibfield  {author} {\bibinfo {author} {\bibfnamefont {P.~J.}\ \bibnamefont
  {Cavalcanti}}, \bibinfo {author} {\bibfnamefont {J.~H.}\ \bibnamefont
  {Selby}}, \bibinfo {author} {\bibfnamefont {J.}~\bibnamefont {Sikora}},
  \bibinfo {author} {\bibfnamefont {T.~D.}\ \bibnamefont {Galley}},\ and\
  \bibinfo {author} {\bibfnamefont {A.~B.}\ \bibnamefont {Sainz}},\ }\href
  {https://doi.org/10.1038/s41534-022-00574-8} {\bibfield  {journal} {\bibinfo
  {journal} {npj Quantum Information}\ }\textbf {\bibinfo {volume} {8}},\
  \bibinfo {pages} {1} (\bibinfo {year} {2022})}\BibitemShut {NoStop}%
\bibitem [{\citenamefont {Wood}\ and\ \citenamefont
  {Spekkens}(2015)}]{wood2015lesson}%
  \BibitemOpen
  \bibfield  {author} {\bibinfo {author} {\bibfnamefont {C.~J.}\ \bibnamefont
  {Wood}}\ and\ \bibinfo {author} {\bibfnamefont {R.~W.}\ \bibnamefont
  {Spekkens}},\ }\href {https://dx.doi.org/10.1088/1367-2630/17/3/033002}
  {\bibfield  {journal} {\bibinfo  {journal} {New J. Phys.}\ }\textbf {\bibinfo
  {volume} {17}},\ \bibinfo {pages} {033002} (\bibinfo {year}
  {2015})}\BibitemShut {NoStop}%
\bibitem [{\citenamefont {Daley}\ \emph {et~al.}(2022)\citenamefont {Daley},
  \citenamefont {Resch},\ and\ \citenamefont
  {Spekkens}}]{daley2021experimentally}%
  \BibitemOpen
  \bibfield  {author} {\bibinfo {author} {\bibfnamefont {P.~J.}\ \bibnamefont
  {Daley}}, \bibinfo {author} {\bibfnamefont {K.~J.}\ \bibnamefont {Resch}},\
  and\ \bibinfo {author} {\bibfnamefont {R.~W.}\ \bibnamefont {Spekkens}},\
  }\href {https://doi.org/10.1103/PhysRevA.105.042220} {\bibfield  {journal}
  {\bibinfo  {journal} {Phys. Rev. A}\ }\textbf {\bibinfo {volume} {105}},\
  \bibinfo {pages} {042220} (\bibinfo {year} {2022})}\BibitemShut {NoStop}%
\bibitem [{\citenamefont {Fine}(1982)}]{Fine}%
  \BibitemOpen
  \bibfield  {author} {\bibinfo {author} {\bibfnamefont {A.}~\bibnamefont
  {Fine}},\ }\href {https://doi.org/10.1103/PhysRevLett.48.291} {\bibfield
  {journal} {\bibinfo  {journal} {Phys. Rev. Lett.}\ }\textbf {\bibinfo
  {volume} {48}},\ \bibinfo {pages} {291} (\bibinfo {year} {1982})}\BibitemShut
  {NoStop}%
\bibitem [{\citenamefont {Barrett}\ \emph {et~al.}(2002)\citenamefont
  {Barrett}, \citenamefont {Collins}, \citenamefont {Hardy}, \citenamefont
  {Kent},\ and\ \citenamefont {Popescu}}]{barrett2002quantum}%
  \BibitemOpen
  \bibfield  {author} {\bibinfo {author} {\bibfnamefont {J.}~\bibnamefont
  {Barrett}}, \bibinfo {author} {\bibfnamefont {D.}~\bibnamefont {Collins}},
  \bibinfo {author} {\bibfnamefont {L.}~\bibnamefont {Hardy}}, \bibinfo
  {author} {\bibfnamefont {A.}~\bibnamefont {Kent}},\ and\ \bibinfo {author}
  {\bibfnamefont {S.}~\bibnamefont {Popescu}},\ }\href@noop {} {\bibfield
  {journal} {\bibinfo  {journal} {Physical Review A}\ }\textbf {\bibinfo
  {volume} {66}},\ \bibinfo {pages} {042111} (\bibinfo {year}
  {2002})}\BibitemShut {NoStop}%
\bibitem [{\citenamefont {Bierhorst}(2015)}]{bierhorst2015robust}%
  \BibitemOpen
  \bibfield  {author} {\bibinfo {author} {\bibfnamefont {P.}~\bibnamefont
  {Bierhorst}},\ }\href@noop {} {\bibfield  {journal} {\bibinfo  {journal}
  {Journal of Physics A: Mathematical and Theoretical}\ }\textbf {\bibinfo
  {volume} {48}},\ \bibinfo {pages} {195302} (\bibinfo {year}
  {2015})}\BibitemShut {NoStop}%
\bibitem [{\citenamefont {Shalm}\ \emph {et~al.}(2015)\citenamefont {Shalm},
  \citenamefont {Meyer-Scott}, \citenamefont {Christensen}, \citenamefont
  {Bierhorst}, \citenamefont {Wayne}, \citenamefont {Stevens}, \citenamefont
  {Gerrits}, \citenamefont {Glancy}, \citenamefont {Hamel}, \citenamefont
  {Allman} \emph {et~al.}}]{shalm2015strong}%
  \BibitemOpen
  \bibfield  {author} {\bibinfo {author} {\bibfnamefont {L.~K.}\ \bibnamefont
  {Shalm}}, \bibinfo {author} {\bibfnamefont {E.}~\bibnamefont {Meyer-Scott}},
  \bibinfo {author} {\bibfnamefont {B.~G.}\ \bibnamefont {Christensen}},
  \bibinfo {author} {\bibfnamefont {P.}~\bibnamefont {Bierhorst}}, \bibinfo
  {author} {\bibfnamefont {M.~A.}\ \bibnamefont {Wayne}}, \bibinfo {author}
  {\bibfnamefont {M.~J.}\ \bibnamefont {Stevens}}, \bibinfo {author}
  {\bibfnamefont {T.}~\bibnamefont {Gerrits}}, \bibinfo {author} {\bibfnamefont
  {S.}~\bibnamefont {Glancy}}, \bibinfo {author} {\bibfnamefont {D.~R.}\
  \bibnamefont {Hamel}}, \bibinfo {author} {\bibfnamefont {M.~S.}\ \bibnamefont
  {Allman}}, \emph {et~al.},\ }\href@noop {} {\bibfield  {journal} {\bibinfo
  {journal} {Physical review letters}\ }\textbf {\bibinfo {volume} {115}},\
  \bibinfo {pages} {250402} (\bibinfo {year} {2015})}\BibitemShut {NoStop}%
\bibitem [{\citenamefont {Selby}\ \emph {et~al.}()\citenamefont {Selby},
  \citenamefont {Schmid}, \citenamefont {Rossi},\ and\ \citenamefont
  {Sainz}}]{NCsubsystems}%
  \BibitemOpen
  \bibfield  {author} {\bibinfo {author} {\bibfnamefont {J.~H.}\ \bibnamefont
  {Selby}}, \bibinfo {author} {\bibfnamefont {D.}~\bibnamefont {Schmid}},
  \bibinfo {author} {\bibfnamefont {V.}~\bibnamefont {Rossi}},\ and\ \bibinfo
  {author} {\bibfnamefont {A.~B.}\ \bibnamefont {Sainz}},\ }\href@noop {}
  {\enquote {\bibinfo {title} {When failures of tomographic completeness are
  not problematic for noncontextuality},}\ }\bibinfo {note}
  {Forthcoming.}\BibitemShut {Stop}%
\bibitem [{\citenamefont {Gottesman}(1998)}]{gottesman1998heisenberg}%
  \BibitemOpen
  \bibfield  {author} {\bibinfo {author} {\bibfnamefont {D.}~\bibnamefont
  {Gottesman}},\ }\href {https://arxiv.org/abs/quant-ph/9807006} {\  (\bibinfo
  {year} {1998})},\ \bibinfo {note} {quant-ph/9807006}\BibitemShut {NoStop}%
\bibitem [{\citenamefont {Schmid}\ \emph {et~al.}(2022)\citenamefont {Schmid},
  \citenamefont {Du}, \citenamefont {Selby},\ and\ \citenamefont
  {Pusey}}]{Schmid2022Stabilizer}%
  \BibitemOpen
  \bibfield  {author} {\bibinfo {author} {\bibfnamefont {D.}~\bibnamefont
  {Schmid}}, \bibinfo {author} {\bibfnamefont {H.}~\bibnamefont {Du}}, \bibinfo
  {author} {\bibfnamefont {J.~H.}\ \bibnamefont {Selby}},\ and\ \bibinfo
  {author} {\bibfnamefont {M.~F.}\ \bibnamefont {Pusey}},\ }\href
  {https://doi.org/10.1103/PhysRevLett.129.120403} {\bibfield  {journal}
  {\bibinfo  {journal} {Phys. Rev. Lett.}\ }\textbf {\bibinfo {volume} {129}},\
  \bibinfo {pages} {120403} (\bibinfo {year} {2022})}\BibitemShut {NoStop}%
\bibitem [{\citenamefont {Wolfe}\ \emph {et~al.}(2020)\citenamefont {Wolfe},
  \citenamefont {Schmid}, \citenamefont {Sainz}, \citenamefont {Kunjwal},\ and\
  \citenamefont {Spekkens}}]{Wolfe2020quantifyingbell}%
  \BibitemOpen
  \bibfield  {author} {\bibinfo {author} {\bibfnamefont {E.}~\bibnamefont
  {Wolfe}}, \bibinfo {author} {\bibfnamefont {D.}~\bibnamefont {Schmid}},
  \bibinfo {author} {\bibfnamefont {A.~B.}\ \bibnamefont {Sainz}}, \bibinfo
  {author} {\bibfnamefont {R.}~\bibnamefont {Kunjwal}},\ and\ \bibinfo {author}
  {\bibfnamefont {R.~W.}\ \bibnamefont {Spekkens}},\ }\href
  {https://doi.org/10.22331/q-2020-06-08-280} {\bibfield  {journal} {\bibinfo
  {journal} {{Quantum}}\ }\textbf {\bibinfo {volume} {4}},\ \bibinfo {pages}
  {280} (\bibinfo {year} {2020})}\BibitemShut {NoStop}%
\bibitem [{\citenamefont {Schmid}\ \emph
  {et~al.}(2020{\natexlab{b}})\citenamefont {Schmid}, \citenamefont {Rosset},\
  and\ \citenamefont {Buscemi}}]{Schmid2020typeindependent}%
  \BibitemOpen
  \bibfield  {author} {\bibinfo {author} {\bibfnamefont {D.}~\bibnamefont
  {Schmid}}, \bibinfo {author} {\bibfnamefont {D.}~\bibnamefont {Rosset}},\
  and\ \bibinfo {author} {\bibfnamefont {F.}~\bibnamefont {Buscemi}},\ }\href
  {https://doi.org/10.22331/q-2020-04-30-262} {\bibfield  {journal} {\bibinfo
  {journal} {{Quantum}}\ }\textbf {\bibinfo {volume} {4}},\ \bibinfo {pages}
  {262} (\bibinfo {year} {2020}{\natexlab{b}})}\BibitemShut {NoStop}%
\bibitem [{\citenamefont {Schmid}\ \emph {et~al.}(2023)\citenamefont {Schmid},
  \citenamefont {Fraser}, \citenamefont {Kunjwal}, \citenamefont {Sainz},
  \citenamefont {Wolfe},\ and\ \citenamefont
  {Spekkens}}]{Schmid2023understanding}%
  \BibitemOpen
  \bibfield  {author} {\bibinfo {author} {\bibfnamefont {D.}~\bibnamefont
  {Schmid}}, \bibinfo {author} {\bibfnamefont {T.~C.}\ \bibnamefont {Fraser}},
  \bibinfo {author} {\bibfnamefont {R.}~\bibnamefont {Kunjwal}}, \bibinfo
  {author} {\bibfnamefont {A.~B.}\ \bibnamefont {Sainz}}, \bibinfo {author}
  {\bibfnamefont {E.}~\bibnamefont {Wolfe}},\ and\ \bibinfo {author}
  {\bibfnamefont {R.~W.}\ \bibnamefont {Spekkens}},\ }\href
  {https://doi.org/10.22331/q-2023-12-04-1194} {\bibfield  {journal} {\bibinfo
  {journal} {{Quantum}}\ }\textbf {\bibinfo {volume} {7}},\ \bibinfo {pages}
  {1194} (\bibinfo {year} {2023})}\BibitemShut {NoStop}%
\bibitem [{\citenamefont {Anders}\ and\ \citenamefont {Browne}(2009)}]{browne}%
  \BibitemOpen
  \bibfield  {author} {\bibinfo {author} {\bibfnamefont {J.}~\bibnamefont
  {Anders}}\ and\ \bibinfo {author} {\bibfnamefont {D.~E.}\ \bibnamefont
  {Browne}},\ }\href {https://doi.org/10.1103/PhysRevLett.102.050502}
  {\bibfield  {journal} {\bibinfo  {journal} {Phys. Rev. Lett.}\ }\textbf
  {\bibinfo {volume} {102}},\ \bibinfo {pages} {050502} (\bibinfo {year}
  {2009})}\BibitemShut {NoStop}%
\bibitem [{\citenamefont {Spekkens}(2019)}]{RobPIRSA}%
  \BibitemOpen
  \bibfield  {author} {\bibinfo {author} {\bibfnamefont {R.~W.}\ \bibnamefont
  {Spekkens}},\ }\href {https://pirsa.org/19110120} {\bibfield  {journal}
  {\bibinfo  {journal} {{PIRSA:19110120}}\ } (\bibinfo {year}
  {2019})}\BibitemShut {NoStop}%
\bibitem [{\citenamefont {Raussendorf}(2013)}]{comp1}%
  \BibitemOpen
  \bibfield  {author} {\bibinfo {author} {\bibfnamefont {R.}~\bibnamefont
  {Raussendorf}},\ }\href {http://link.aps.org/doi/10.1103/PhysRevA.88.022322}
  {\bibfield  {journal} {\bibinfo  {journal} {Phys. Rev. A}\ }\textbf {\bibinfo
  {volume} {88}},\ \bibinfo {pages} {022322} (\bibinfo {year}
  {2013})}\BibitemShut {NoStop}%
\bibitem [{\citenamefont {Chailloux}\ \emph {et~al.}(2016)\citenamefont
  {Chailloux}, \citenamefont {Kerenidis}, \citenamefont {Kundu},\ and\
  \citenamefont {Sikora}}]{RAC}%
  \BibitemOpen
  \bibfield  {author} {\bibinfo {author} {\bibfnamefont {A.}~\bibnamefont
  {Chailloux}}, \bibinfo {author} {\bibfnamefont {I.}~\bibnamefont
  {Kerenidis}}, \bibinfo {author} {\bibfnamefont {S.}~\bibnamefont {Kundu}},\
  and\ \bibinfo {author} {\bibfnamefont {J.}~\bibnamefont {Sikora}},\ }\href
  {http://stacks.iop.org/1367-2630/18/i=4/a=045003} {\bibfield  {journal}
  {\bibinfo  {journal} {New J. Phys.}\ }\textbf {\bibinfo {volume} {18}},\
  \bibinfo {pages} {045003} (\bibinfo {year} {2016})}\BibitemShut {NoStop}%
\bibitem [{\citenamefont {Ambainis}\ \emph {et~al.}(2019)\citenamefont
  {Ambainis}, \citenamefont {Banik}, \citenamefont {Chaturvedi}, \citenamefont
  {Kravchenko},\ and\ \citenamefont {Rai}}]{RAC2}%
  \BibitemOpen
  \bibfield  {author} {\bibinfo {author} {\bibfnamefont {A.}~\bibnamefont
  {Ambainis}}, \bibinfo {author} {\bibfnamefont {M.}~\bibnamefont {Banik}},
  \bibinfo {author} {\bibfnamefont {A.}~\bibnamefont {Chaturvedi}}, \bibinfo
  {author} {\bibfnamefont {D.}~\bibnamefont {Kravchenko}},\ and\ \bibinfo
  {author} {\bibfnamefont {A.}~\bibnamefont {Rai}},\ }\href
  {https://link.springer.com/article/10.1007/s11128-019-2228-3} {\bibfield
  {journal} {\bibinfo  {journal} {Quantum Inf. Process.}\ }\textbf {\bibinfo
  {volume} {18}},\ \bibinfo {pages} {111} (\bibinfo {year} {2019})}\BibitemShut
  {NoStop}%
\bibitem [{\citenamefont {Catani}\ \emph {et~al.}(2022)\citenamefont {Catani},
  \citenamefont {Leifer}, \citenamefont {Schmid},\ and\ \citenamefont
  {Spekkens}}]{catani2022reply}%
  \BibitemOpen
  \bibfield  {author} {\bibinfo {author} {\bibfnamefont {L.}~\bibnamefont
  {Catani}}, \bibinfo {author} {\bibfnamefont {M.}~\bibnamefont {Leifer}},
  \bibinfo {author} {\bibfnamefont {D.}~\bibnamefont {Schmid}},\ and\ \bibinfo
  {author} {\bibfnamefont {R.~W.}\ \bibnamefont {Spekkens}},\ }\href
  {https://arxiv.org/abs/2207.11791} {\bibfield  {journal} {\bibinfo  {journal}
  {arXiv preprint arXiv:2207.11791}\ } (\bibinfo {year} {2022})}\BibitemShut
  {NoStop}%
\bibitem [{\citenamefont {Pusey}(2015)}]{robust}%
  \BibitemOpen
  \bibfield  {author} {\bibinfo {author} {\bibfnamefont {M.~F.}\ \bibnamefont
  {Pusey}},\ }\href {https://arxiv.org/abs/1506.04178} {\bibfield  {journal}
  {\bibinfo  {journal} {arXiv:1506.04178}\ } (\bibinfo {year}
  {2015})}\BibitemShut {NoStop}%
\bibitem [{\citenamefont {Mazurek}\ \emph {et~al.}(2016)\citenamefont
  {Mazurek}, \citenamefont {Pusey}, \citenamefont {Kunjwal}, \citenamefont
  {Resch},\ and\ \citenamefont {Spekkens}}]{mazurek2016experimental}%
  \BibitemOpen
  \bibfield  {author} {\bibinfo {author} {\bibfnamefont {M.~D.}\ \bibnamefont
  {Mazurek}}, \bibinfo {author} {\bibfnamefont {M.~F.}\ \bibnamefont {Pusey}},
  \bibinfo {author} {\bibfnamefont {R.}~\bibnamefont {Kunjwal}}, \bibinfo
  {author} {\bibfnamefont {K.~J.}\ \bibnamefont {Resch}},\ and\ \bibinfo
  {author} {\bibfnamefont {R.~W.}\ \bibnamefont {Spekkens}},\ }\href
  {https://doi.org/10.1038/ncomms11780} {\bibfield  {journal} {\bibinfo
  {journal} {Nature communications}\ }\textbf {\bibinfo {volume} {7}},\
  \bibinfo {pages} {1} (\bibinfo {year} {2016})}\BibitemShut {NoStop}%
\bibitem [{\citenamefont {Schmid}\ \emph {et~al.}(2018)\citenamefont {Schmid},
  \citenamefont {Spekkens},\ and\ \citenamefont {Wolfe}}]{Schmid2018}%
  \BibitemOpen
  \bibfield  {author} {\bibinfo {author} {\bibfnamefont {D.}~\bibnamefont
  {Schmid}}, \bibinfo {author} {\bibfnamefont {R.~W.}\ \bibnamefont
  {Spekkens}},\ and\ \bibinfo {author} {\bibfnamefont {E.}~\bibnamefont
  {Wolfe}},\ }\href {https://doi.org/10.1103/PhysRevA.97.062103} {\bibfield
  {journal} {\bibinfo  {journal} {Phys. Rev. A}\ }\textbf {\bibinfo {volume}
  {97}},\ \bibinfo {pages} {062103} (\bibinfo {year} {2018})}\BibitemShut
  {NoStop}%
\bibitem [{\citenamefont {Bell}(1966)}]{Bell2}%
  \BibitemOpen
  \bibfield  {author} {\bibinfo {author} {\bibfnamefont {J.~S.}\ \bibnamefont
  {Bell}},\ }\href {https://doi.org/10.1103/RevModPhys.38.447} {\bibfield
  {journal} {\bibinfo  {journal} {Rev. Mod. Phys.}\ }\textbf {\bibinfo {volume}
  {38}},\ \bibinfo {pages} {447} (\bibinfo {year} {1966})}\BibitemShut
  {NoStop}%
\bibitem [{\citenamefont {de~Broglie}(1960)}]{de1960non}%
  \BibitemOpen
  \bibfield  {author} {\bibinfo {author} {\bibfnamefont {L.}~\bibnamefont
  {de~Broglie}},\ }\href {https://books.google.pl/books?id=2wI\_wwEACAAJ}
  {\emph {\bibinfo {title} {Non-linear Wave Mechanics}}}\ (\bibinfo
  {publisher} {Elsevier},\ \bibinfo {year} {1960})\BibitemShut {NoStop}%
\bibitem [{\citenamefont {Bohm}(1952{\natexlab{a}})}]{Bohm}%
  \BibitemOpen
  \bibfield  {author} {\bibinfo {author} {\bibfnamefont {D.}~\bibnamefont
  {Bohm}},\ }\href {https://doi.org/10.1103/PhysRev.85.166} {\bibfield
  {journal} {\bibinfo  {journal} {Phys. Rev.}\ }\textbf {\bibinfo {volume}
  {85}},\ \bibinfo {pages} {166} (\bibinfo {year}
  {1952}{\natexlab{a}})}\BibitemShut {NoStop}%
\bibitem [{\citenamefont {Bohm}(1952{\natexlab{b}})}]{Bohm2}%
  \BibitemOpen
  \bibfield  {author} {\bibinfo {author} {\bibfnamefont {D.}~\bibnamefont
  {Bohm}},\ }\href {https://doi.org/10.1103/PhysRev.85.180} {\bibfield
  {journal} {\bibinfo  {journal} {Phys. Rev.}\ }\textbf {\bibinfo {volume}
  {85}},\ \bibinfo {pages} {180} (\bibinfo {year}
  {1952}{\natexlab{b}})}\BibitemShut {NoStop}%
\bibitem [{\citenamefont {Spekkens}(2007)}]{spekkens2007evidence}%
  \BibitemOpen
  \bibfield  {author} {\bibinfo {author} {\bibfnamefont {R.~W.}\ \bibnamefont
  {Spekkens}},\ }\href {http://link.aps.org/doi/10.1103/PhysRevA.75.032110}
  {\bibfield  {journal} {\bibinfo  {journal} {Phys. Rev. A}\ }\textbf {\bibinfo
  {volume} {75}},\ \bibinfo {pages} {032110} (\bibinfo {year}
  {2007})}\BibitemShut {NoStop}%
\bibitem [{\citenamefont {Bartlett}\ \emph {et~al.}(2012)\citenamefont
  {Bartlett}, \citenamefont {Rudolph},\ and\ \citenamefont
  {Spekkens}}]{bartlett2012reconstruction}%
  \BibitemOpen
  \bibfield  {author} {\bibinfo {author} {\bibfnamefont {S.~D.}\ \bibnamefont
  {Bartlett}}, \bibinfo {author} {\bibfnamefont {T.}~\bibnamefont {Rudolph}},\
  and\ \bibinfo {author} {\bibfnamefont {R.~W.}\ \bibnamefont {Spekkens}},\
  }\href {https://journals.aps.org/pra/abstract/10.1103/PhysRevA.86.012103}
  {\bibfield  {journal} {\bibinfo  {journal} {Phys. Rev. A}\ }\textbf {\bibinfo
  {volume} {86}},\ \bibinfo {pages} {012103} (\bibinfo {year}
  {2012})}\BibitemShut {NoStop}%
\bibitem [{\citenamefont {Catani}\ \emph {et~al.}(2023)\citenamefont {Catani},
  \citenamefont {Leifer}, \citenamefont {Schmid},\ and\ \citenamefont
  {Spekkens}}]{Catani2023whyinterference}%
  \BibitemOpen
  \bibfield  {author} {\bibinfo {author} {\bibfnamefont {L.}~\bibnamefont
  {Catani}}, \bibinfo {author} {\bibfnamefont {M.}~\bibnamefont {Leifer}},
  \bibinfo {author} {\bibfnamefont {D.}~\bibnamefont {Schmid}},\ and\ \bibinfo
  {author} {\bibfnamefont {R.~W.}\ \bibnamefont {Spekkens}},\ }\href
  {https://doi.org/10.22331/q-2023-09-25-1119} {\bibfield  {journal} {\bibinfo
  {journal} {{Quantum}}\ }\textbf {\bibinfo {volume} {7}},\ \bibinfo {pages}
  {1119} (\bibinfo {year} {2023})}\BibitemShut {NoStop}%
\bibitem [{\citenamefont {Spekkens}(2016)}]{epistricted}%
  \BibitemOpen
  \bibfield  {author} {\bibinfo {author} {\bibfnamefont {R.~W.}\ \bibnamefont
  {Spekkens}},\ }\enquote {\bibinfo {title} {{Quasi-Quantization: Classical
  Statistical Theories with an Epistemic Restriction}},}\ in\ \href
  {https://doi.org/10.1007/978-94-017-7303-4_4} {\emph {\bibinfo {booktitle}
  {{Quantum Theory: Informational Foundations and Foils}}}},\ \bibinfo {editor}
  {edited by\ \bibinfo {editor} {\bibfnamefont {G.}~\bibnamefont {Chiribella}}\
  and\ \bibinfo {editor} {\bibfnamefont {R.~W.}\ \bibnamefont {Spekkens}}}\
  (\bibinfo  {publisher} {Springer Netherlands},\ \bibinfo {address}
  {Dordrecht},\ \bibinfo {year} {2016})\ pp.\ \bibinfo {pages}
  {83--135}\BibitemShut {NoStop}%
\bibitem [{\citenamefont {{Pusey}}\ \emph {et~al.}(2019)\citenamefont
  {{Pusey}}, \citenamefont {{del Rio}},\ and\ \citenamefont
  {{Meyer}}}]{PuseydelRio}%
  \BibitemOpen
  \bibfield  {author} {\bibinfo {author} {\bibfnamefont {M.~F.}\ \bibnamefont
  {{Pusey}}}, \bibinfo {author} {\bibfnamefont {L.}~\bibnamefont {{del Rio}}},\
  and\ \bibinfo {author} {\bibfnamefont {B.}~\bibnamefont {{Meyer}}},\ }\href
  {https://arxiv.org/abs/1904.08699} {\bibfield  {journal} {\bibinfo  {journal}
  {arXiv:1904.08699}\ } (\bibinfo {year} {2019})}\BibitemShut {NoStop}%
\bibitem [{\citenamefont {M\"uller}\ and\ \citenamefont
  {Garner}(2023)}]{muller2023testing}%
  \BibitemOpen
  \bibfield  {author} {\bibinfo {author} {\bibfnamefont {M.~P.}\ \bibnamefont
  {M\"uller}}\ and\ \bibinfo {author} {\bibfnamefont {A.~J.~P.}\ \bibnamefont
  {Garner}},\ }\href {https://doi.org/10.1103/PhysRevX.13.041001} {\bibfield
  {journal} {\bibinfo  {journal} {Phys. Rev. X}\ }\textbf {\bibinfo {volume}
  {13}},\ \bibinfo {pages} {041001} (\bibinfo {year} {2023})}\BibitemShut
  {NoStop}%
\bibitem [{\citenamefont {Chiribella}(2018)}]{Giuliosubsystems}%
  \BibitemOpen
  \bibfield  {author} {\bibinfo {author} {\bibfnamefont {G.}~\bibnamefont
  {Chiribella}},\ }\bibfield  {journal} {\bibinfo  {journal} {Entropy}\
  }\textbf {\bibinfo {volume} {20}},\ \href {https://doi.org/10.3390/e20050358}
  {10.3390/e20050358} (\bibinfo {year} {2018})\BibitemShut {NoStop}%
\bibitem [{\citenamefont {Kr{\"a}mer}\ and\ \citenamefont
  {Del~Rio}(2018)}]{kramer2018operational}%
  \BibitemOpen
  \bibfield  {author} {\bibinfo {author} {\bibfnamefont {L.}~\bibnamefont
  {Kr{\"a}mer}}\ and\ \bibinfo {author} {\bibfnamefont {L.}~\bibnamefont
  {Del~Rio}},\ }\href {https://doi.org/10.1098/rsta.2017.0321} {\bibfield
  {journal} {\bibinfo  {journal} {Philosophical Transactions of the Royal
  Society A: Mathematical, Physical and Engineering Sciences}\ }\textbf
  {\bibinfo {volume} {376}},\ \bibinfo {pages} {20170321} (\bibinfo {year}
  {2018})}\BibitemShut {NoStop}%
\bibitem [{\citenamefont {Pearl}(2009)}]{pearl2009causality}%
  \BibitemOpen
  \bibfield  {author} {\bibinfo {author} {\bibfnamefont {J.}~\bibnamefont
  {Pearl}},\ }\href {https://doi.org/10.1017/CBO9780511803161} {\emph {\bibinfo
  {title} {Causality}}}\ (\bibinfo  {publisher} {Cambridge university press},\
  \bibinfo {year} {2009})\BibitemShut {NoStop}%
\bibitem [{\citenamefont {Spirtes}\ \emph {et~al.}(2000)\citenamefont
  {Spirtes}, \citenamefont {Glymour}, \citenamefont {Scheines},\ and\
  \citenamefont {Heckerman}}]{spirtes2000causation}%
  \BibitemOpen
  \bibfield  {author} {\bibinfo {author} {\bibfnamefont {P.}~\bibnamefont
  {Spirtes}}, \bibinfo {author} {\bibfnamefont {C.~N.}\ \bibnamefont
  {Glymour}}, \bibinfo {author} {\bibfnamefont {R.}~\bibnamefont {Scheines}},\
  and\ \bibinfo {author} {\bibfnamefont {D.}~\bibnamefont {Heckerman}},\ }\href
  {https://mitpress.mit.edu/books/causation-prediction-and-search-second-edition}
  {\emph {\bibinfo {title} {Causation, prediction, and search}}}\ (\bibinfo
  {publisher} {MIT press},\ \bibinfo {year} {2000})\BibitemShut {NoStop}%
\bibitem [{\citenamefont {Costa}\ and\ \citenamefont
  {Shrapnel}(2016)}]{costa2016quantum}%
  \BibitemOpen
  \bibfield  {author} {\bibinfo {author} {\bibfnamefont {F.}~\bibnamefont
  {Costa}}\ and\ \bibinfo {author} {\bibfnamefont {S.}~\bibnamefont
  {Shrapnel}},\ }\href
  {https://iopscience.iop.org/article/10.1088/1367-2630/18/6/063032} {\bibfield
   {journal} {\bibinfo  {journal} {New J. Phys.}\ }\textbf {\bibinfo {volume}
  {18}},\ \bibinfo {pages} {063032} (\bibinfo {year} {2016})}\BibitemShut
  {NoStop}%
\bibitem [{\citenamefont {Allen}\ \emph {et~al.}(2017)\citenamefont {Allen},
  \citenamefont {Barrett}, \citenamefont {Horsman}, \citenamefont {Lee},\ and\
  \citenamefont {Spekkens}}]{allen2017quantum}%
  \BibitemOpen
  \bibfield  {author} {\bibinfo {author} {\bibfnamefont {J.-M.~A.}\
  \bibnamefont {Allen}}, \bibinfo {author} {\bibfnamefont {J.}~\bibnamefont
  {Barrett}}, \bibinfo {author} {\bibfnamefont {D.~C.}\ \bibnamefont
  {Horsman}}, \bibinfo {author} {\bibfnamefont {C.~M.}\ \bibnamefont {Lee}},\
  and\ \bibinfo {author} {\bibfnamefont {R.~W.}\ \bibnamefont {Spekkens}},\
  }\href {https://doi.org/10.1103/PhysRevX.7.031021} {\bibfield  {journal}
  {\bibinfo  {journal} {Phys. Rev. X}\ }\textbf {\bibinfo {volume} {7}},\
  \bibinfo {pages} {031021} (\bibinfo {year} {2017})}\BibitemShut {NoStop}%
\bibitem [{\citenamefont {{Barrett}}\ \emph {et~al.}(2019)\citenamefont
  {{Barrett}}, \citenamefont {{Lorenz}},\ and\ \citenamefont
  {{Oreshkov}}}]{Barrett2019}%
  \BibitemOpen
  \bibfield  {author} {\bibinfo {author} {\bibfnamefont {J.}~\bibnamefont
  {{Barrett}}}, \bibinfo {author} {\bibfnamefont {R.}~\bibnamefont
  {{Lorenz}}},\ and\ \bibinfo {author} {\bibfnamefont {O.}~\bibnamefont
  {{Oreshkov}}},\ }\href
  {https://ui.adsabs.harvard.edu/abs/2019arXiv190610726B} {\  (\bibinfo {year}
  {2019})},\ \Eprint {https://arxiv.org/abs/1906.10726} {arXiv:1906.10726}
  \BibitemShut {NoStop}%
\bibitem [{\citenamefont {Spekkens}(2008)}]{negativity}%
  \BibitemOpen
  \bibfield  {author} {\bibinfo {author} {\bibfnamefont {R.~W.}\ \bibnamefont
  {Spekkens}},\ }\href {http://link.aps.org/doi/10.1103/PhysRevLett.101.020401}
  {\bibfield  {journal} {\bibinfo  {journal} {Phys. Rev. Lett.}\ }\textbf
  {\bibinfo {volume} {101}},\ \bibinfo {pages} {020401} (\bibinfo {year}
  {2008})}\BibitemShut {NoStop}%
\bibitem [{\citenamefont {Lillystone}\ \emph {et~al.}(2019)\citenamefont
  {Lillystone}, \citenamefont {Wallman},\ and\ \citenamefont
  {Emerson}}]{Lillystone2019}%
  \BibitemOpen
  \bibfield  {author} {\bibinfo {author} {\bibfnamefont {P.}~\bibnamefont
  {Lillystone}}, \bibinfo {author} {\bibfnamefont {J.~J.}\ \bibnamefont
  {Wallman}},\ and\ \bibinfo {author} {\bibfnamefont {J.}~\bibnamefont
  {Emerson}},\ }\href {https://doi.org/10.1103/PhysRevLett.122.140405}
  {\bibfield  {journal} {\bibinfo  {journal} {Phys. Rev. Lett.}\ }\textbf
  {\bibinfo {volume} {122}},\ \bibinfo {pages} {140405} (\bibinfo {year}
  {2019})}\BibitemShut {NoStop}%
\bibitem [{\citenamefont {Fritz}(2012)}]{FritzBeyondBellI}%
  \BibitemOpen
  \bibfield  {author} {\bibinfo {author} {\bibfnamefont {T.}~\bibnamefont
  {Fritz}},\ }\href {http://stacks.iop.org/1367-2630/14/i=10/a=103001}
  {\bibfield  {journal} {\bibinfo  {journal} {New J. Physics}\ }\textbf
  {\bibinfo {volume} {14}},\ \bibinfo {pages} {103001} (\bibinfo {year}
  {2012})}\BibitemShut {NoStop}%
\bibitem [{\citenamefont {Tavakoli}\ \emph {et~al.}(2021)\citenamefont
  {Tavakoli}, \citenamefont {Pozas-Kerstjens}, \citenamefont {Renou} \emph
  {et~al.}}]{tavakoli2021bell}%
  \BibitemOpen
  \bibfield  {author} {\bibinfo {author} {\bibfnamefont {A.}~\bibnamefont
  {Tavakoli}}, \bibinfo {author} {\bibfnamefont {A.}~\bibnamefont
  {Pozas-Kerstjens}}, \bibinfo {author} {\bibfnamefont {M.-O.}\ \bibnamefont
  {Renou}}, \emph {et~al.},\ }\href {https://doi.org/10.1088/1361-6633/ac41bb}
  {\bibfield  {journal} {\bibinfo  {journal} {Reports on Progress in Physics}\
  } (\bibinfo {year} {2021})}\BibitemShut {NoStop}%
\bibitem [{\citenamefont {Chaves}\ \emph {et~al.}(2021)\citenamefont {Chaves},
  \citenamefont {Moreno}, \citenamefont {Polino}, \citenamefont {Poderini},
  \citenamefont {Agresti}, \citenamefont {Suprano}, \citenamefont {Barros},
  \citenamefont {Carvacho}, \citenamefont {Wolfe}, \citenamefont {Canabarro},
  \citenamefont {Spekkens},\ and\ \citenamefont
  {Sciarrino}}]{ChavesNetworks2021}%
  \BibitemOpen
  \bibfield  {author} {\bibinfo {author} {\bibfnamefont {R.}~\bibnamefont
  {Chaves}}, \bibinfo {author} {\bibfnamefont {G.}~\bibnamefont {Moreno}},
  \bibinfo {author} {\bibfnamefont {E.}~\bibnamefont {Polino}}, \bibinfo
  {author} {\bibfnamefont {D.}~\bibnamefont {Poderini}}, \bibinfo {author}
  {\bibfnamefont {I.}~\bibnamefont {Agresti}}, \bibinfo {author} {\bibfnamefont
  {A.}~\bibnamefont {Suprano}}, \bibinfo {author} {\bibfnamefont {M.~R.}\
  \bibnamefont {Barros}}, \bibinfo {author} {\bibfnamefont {G.}~\bibnamefont
  {Carvacho}}, \bibinfo {author} {\bibfnamefont {E.}~\bibnamefont {Wolfe}},
  \bibinfo {author} {\bibfnamefont {A.}~\bibnamefont {Canabarro}}, \bibinfo
  {author} {\bibfnamefont {R.~W.}\ \bibnamefont {Spekkens}},\ and\ \bibinfo
  {author} {\bibfnamefont {F.}~\bibnamefont {Sciarrino}},\ }\href
  {https://doi.org/10.1103/PRXQuantum.2.040323} {\bibfield  {journal} {\bibinfo
   {journal} {PRX Quantum}\ }\textbf {\bibinfo {volume} {2}},\ \bibinfo {pages}
  {040323} (\bibinfo {year} {2021})}\BibitemShut {NoStop}%
\bibitem [{\citenamefont {Schmid}\ \emph {et~al.}()\citenamefont {Schmid},
  \citenamefont {Selby},\ and\ \citenamefont {Spekkens}}]{fromBelltoNC}%
  \BibitemOpen
  \bibfield  {author} {\bibinfo {author} {\bibfnamefont {D.}~\bibnamefont
  {Schmid}}, \bibinfo {author} {\bibfnamefont {J.}~\bibnamefont {Selby}},\ and\
  \bibinfo {author} {\bibfnamefont {R.~W.}\ \bibnamefont {Spekkens}},\
  }\bibinfo {note} {forthcoming}\BibitemShut {NoStop}%
\bibitem [{\citenamefont {Leifer}\ and\ \citenamefont {Spekkens}(2005)}]{PP1}%
  \BibitemOpen
  \bibfield  {author} {\bibinfo {author} {\bibfnamefont {M.~S.}\ \bibnamefont
  {Leifer}}\ and\ \bibinfo {author} {\bibfnamefont {R.~W.}\ \bibnamefont
  {Spekkens}},\ }\href {http://link.aps.org/doi/10.1103/PhysRevLett.95.200405}
  {\bibfield  {journal} {\bibinfo  {journal} {Phys. Rev. Lett.}\ }\textbf
  {\bibinfo {volume} {95}},\ \bibinfo {pages} {200405} (\bibinfo {year}
  {2005})}\BibitemShut {NoStop}%
\bibitem [{\citenamefont {Pusey}\ and\ \citenamefont {Leifer}(2015)}]{PP2}%
  \BibitemOpen
  \bibfield  {author} {\bibinfo {author} {\bibfnamefont {M.~F.}\ \bibnamefont
  {Pusey}}\ and\ \bibinfo {author} {\bibfnamefont {M.~S.}\ \bibnamefont
  {Leifer}}\ }(\bibinfo  {publisher} {EPTCS},\ \bibinfo {year} {2015})\ pp.\
  \bibinfo {pages} {295--306}\BibitemShut {NoStop}%
\bibitem [{\citenamefont {Spekkens}\ \emph {et~al.}(2009)\citenamefont
  {Spekkens}, \citenamefont {Buzacott}, \citenamefont {Keehn}, \citenamefont
  {Toner},\ and\ \citenamefont {Pryde}}]{POM}%
  \BibitemOpen
  \bibfield  {author} {\bibinfo {author} {\bibfnamefont {R.~W.}\ \bibnamefont
  {Spekkens}}, \bibinfo {author} {\bibfnamefont {D.~H.}\ \bibnamefont
  {Buzacott}}, \bibinfo {author} {\bibfnamefont {A.~J.}\ \bibnamefont {Keehn}},
  \bibinfo {author} {\bibfnamefont {B.}~\bibnamefont {Toner}},\ and\ \bibinfo
  {author} {\bibfnamefont {G.~J.}\ \bibnamefont {Pryde}},\ }\href
  {https://doi.org/10.1103/PhysRevLett.102.010401} {\bibfield  {journal}
  {\bibinfo  {journal} {Phys. Rev. Lett.}\ }\textbf {\bibinfo {volume} {102}},\
  \bibinfo {pages} {010401} (\bibinfo {year} {2009})}\BibitemShut {NoStop}%
\bibitem [{\citenamefont {Liang}\ \emph {et~al.}(2011)\citenamefont {Liang},
  \citenamefont {Spekkens},\ and\ \citenamefont {Wiseman}}]{parable}%
  \BibitemOpen
  \bibfield  {author} {\bibinfo {author} {\bibfnamefont {Y.-C.}\ \bibnamefont
  {Liang}}, \bibinfo {author} {\bibfnamefont {R.~W.}\ \bibnamefont
  {Spekkens}},\ and\ \bibinfo {author} {\bibfnamefont {H.~M.}\ \bibnamefont
  {Wiseman}},\ }\href
  {http://www.sciencedirect.com/science/article/pii/S0370157311001517}
  {\bibfield  {journal} {\bibinfo  {journal} {Physics Reports}\ }\textbf
  {\bibinfo {volume} {506}},\ \bibinfo {pages} {1} (\bibinfo {year}
  {2011})}\BibitemShut {NoStop}%
\bibitem [{\citenamefont {Pusey}(2014)}]{AWV}%
  \BibitemOpen
  \bibfield  {author} {\bibinfo {author} {\bibfnamefont {M.~F.}\ \bibnamefont
  {Pusey}},\ }\href {http://link.aps.org/doi/10.1103/PhysRevLett.113.200401}
  {\bibfield  {journal} {\bibinfo  {journal} {Phys. Rev. Lett.}\ }\textbf
  {\bibinfo {volume} {113}},\ \bibinfo {pages} {200401} (\bibinfo {year}
  {2014})}\BibitemShut {NoStop}%
\bibitem [{\citenamefont {Kunjwal}\ and\ \citenamefont
  {Spekkens}(2015)}]{operationalks}%
  \BibitemOpen
  \bibfield  {author} {\bibinfo {author} {\bibfnamefont {R.}~\bibnamefont
  {Kunjwal}}\ and\ \bibinfo {author} {\bibfnamefont {R.~W.}\ \bibnamefont
  {Spekkens}},\ }\href {https://doi.org/10.1103/PhysRevLett.115.110403}
  {\bibfield  {journal} {\bibinfo  {journal} {Phys. Rev. Lett.}\ }\textbf
  {\bibinfo {volume} {115}},\ \bibinfo {pages} {110403} (\bibinfo {year}
  {2015})}\BibitemShut {NoStop}%
\bibitem [{\citenamefont {Schmid}\ and\ \citenamefont
  {Spekkens}(2018)}]{schmid2018contextual}%
  \BibitemOpen
  \bibfield  {author} {\bibinfo {author} {\bibfnamefont {D.}~\bibnamefont
  {Schmid}}\ and\ \bibinfo {author} {\bibfnamefont {R.~W.}\ \bibnamefont
  {Spekkens}},\ }\href
  {https://journals.aps.org/prx/abstract/10.1103/PhysRevX.8.011015} {\bibfield
  {journal} {\bibinfo  {journal} {Phys. Rev. X}\ }\textbf {\bibinfo {volume}
  {8}},\ \bibinfo {pages} {011015} (\bibinfo {year} {2018})}\BibitemShut
  {NoStop}%
\bibitem [{\citenamefont {Kunjwal}\ \emph {et~al.}(2019)\citenamefont
  {Kunjwal}, \citenamefont {Lostaglio},\ and\ \citenamefont {Pusey}}]{KLP19}%
  \BibitemOpen
  \bibfield  {author} {\bibinfo {author} {\bibfnamefont {R.}~\bibnamefont
  {Kunjwal}}, \bibinfo {author} {\bibfnamefont {M.}~\bibnamefont {Lostaglio}},\
  and\ \bibinfo {author} {\bibfnamefont {M.~F.}\ \bibnamefont {Pusey}},\ }\href
  {https://doi.org/10.1103/PhysRevA.100.042116} {\bibfield  {journal} {\bibinfo
   {journal} {Phys. Rev. A}\ }\textbf {\bibinfo {volume} {100}},\ \bibinfo
  {pages} {042116} (\bibinfo {year} {2019})}\BibitemShut {NoStop}%
\bibitem [{\citenamefont {Saha}\ and\ \citenamefont
  {Chaturvedi}(2019)}]{saha2019preparation}%
  \BibitemOpen
  \bibfield  {author} {\bibinfo {author} {\bibfnamefont {D.}~\bibnamefont
  {Saha}}\ and\ \bibinfo {author} {\bibfnamefont {A.}~\bibnamefont
  {Chaturvedi}},\ }\href {https://doi.org/10.1103/PhysRevA.100.022108}
  {\bibfield  {journal} {\bibinfo  {journal} {Phys. Rev. A}\ }\textbf {\bibinfo
  {volume} {100}},\ \bibinfo {pages} {022108} (\bibinfo {year}
  {2019})}\BibitemShut {NoStop}%
\bibitem [{\citenamefont {Lostaglio}(2020)}]{contextmetrology}%
  \BibitemOpen
  \bibfield  {author} {\bibinfo {author} {\bibfnamefont {M.}~\bibnamefont
  {Lostaglio}},\ }\href {https://doi.org/10.1103/PhysRevLett.125.230603}
  {\bibfield  {journal} {\bibinfo  {journal} {Phys. Rev. Lett.}\ }\textbf
  {\bibinfo {volume} {125}},\ \bibinfo {pages} {230603} (\bibinfo {year}
  {2020})}\BibitemShut {NoStop}%
\bibitem [{\citenamefont {Lostaglio}\ and\ \citenamefont
  {Senno}(2020)}]{cloningcontext}%
  \BibitemOpen
  \bibfield  {author} {\bibinfo {author} {\bibfnamefont {M.}~\bibnamefont
  {Lostaglio}}\ and\ \bibinfo {author} {\bibfnamefont {G.}~\bibnamefont
  {Senno}},\ }\href {https://doi.org/10.22331/q-2020-04-27-258} {\bibfield
  {journal} {\bibinfo  {journal} {Quantum}\ }\textbf {\bibinfo {volume} {4}},\
  \bibinfo {pages} {258} (\bibinfo {year} {2020})}\BibitemShut {NoStop}%
\bibitem [{\citenamefont {Yadavalli}\ and\ \citenamefont
  {Kunjwal}(2022)}]{Yadavalli2020}%
  \BibitemOpen
  \bibfield  {author} {\bibinfo {author} {\bibfnamefont {S.~A.}\ \bibnamefont
  {Yadavalli}}\ and\ \bibinfo {author} {\bibfnamefont {R.}~\bibnamefont
  {Kunjwal}},\ }\href {https://doi.org/10.22331/q-2022-10-13-839} {\bibfield
  {journal} {\bibinfo  {journal} {{Quantum}}\ }\textbf {\bibinfo {volume}
  {6}},\ \bibinfo {pages} {839} (\bibinfo {year} {2022})}\BibitemShut {NoStop}%
\bibitem [{\citenamefont {Selby}\ \emph
  {et~al.}(2023{\natexlab{b}})\citenamefont {Selby}, \citenamefont {Schmid},
  \citenamefont {Wolfe}, \citenamefont {Sainz}, \citenamefont {Kunjwal},\ and\
  \citenamefont {Spekkens}}]{selby2023incompatibility}%
  \BibitemOpen
  \bibfield  {author} {\bibinfo {author} {\bibfnamefont {J.~H.}\ \bibnamefont
  {Selby}}, \bibinfo {author} {\bibfnamefont {D.}~\bibnamefont {Schmid}},
  \bibinfo {author} {\bibfnamefont {E.}~\bibnamefont {Wolfe}}, \bibinfo
  {author} {\bibfnamefont {A.~B.}\ \bibnamefont {Sainz}}, \bibinfo {author}
  {\bibfnamefont {R.}~\bibnamefont {Kunjwal}},\ and\ \bibinfo {author}
  {\bibfnamefont {R.~W.}\ \bibnamefont {Spekkens}},\ }\href
  {https://doi.org/10.1103/PhysRevLett.130.230201} {\bibfield  {journal}
  {\bibinfo  {journal} {Phys. Rev. Lett.}\ }\textbf {\bibinfo {volume} {130}},\
  \bibinfo {pages} {230201} (\bibinfo {year} {2023}{\natexlab{b}})}\BibitemShut
  {NoStop}%
\bibitem [{\citenamefont {Roch~i Carceller}\ \emph {et~al.}(2022)\citenamefont
  {Roch~i Carceller}, \citenamefont {Flatt}, \citenamefont {Lee}, \citenamefont
  {Bae},\ and\ \citenamefont {Brask}}]{Roch2021}%
  \BibitemOpen
  \bibfield  {author} {\bibinfo {author} {\bibfnamefont {C.}~\bibnamefont
  {Roch~i Carceller}}, \bibinfo {author} {\bibfnamefont {K.}~\bibnamefont
  {Flatt}}, \bibinfo {author} {\bibfnamefont {H.}~\bibnamefont {Lee}}, \bibinfo
  {author} {\bibfnamefont {J.}~\bibnamefont {Bae}},\ and\ \bibinfo {author}
  {\bibfnamefont {J.~B.}\ \bibnamefont {Brask}},\ }\href
  {https://doi.org/10.1103/PhysRevLett.129.050501} {\bibfield  {journal}
  {\bibinfo  {journal} {Phys. Rev. Lett.}\ }\textbf {\bibinfo {volume} {129}},\
  \bibinfo {pages} {050501} (\bibinfo {year} {2022})}\BibitemShut {NoStop}%
\bibitem [{\citenamefont {Flatt}\ \emph {et~al.}(2022)\citenamefont {Flatt},
  \citenamefont {Lee}, \citenamefont {Carceller}, \citenamefont {Brask},\ and\
  \citenamefont {Bae}}]{Flatt2021}%
  \BibitemOpen
  \bibfield  {author} {\bibinfo {author} {\bibfnamefont {K.}~\bibnamefont
  {Flatt}}, \bibinfo {author} {\bibfnamefont {H.}~\bibnamefont {Lee}}, \bibinfo
  {author} {\bibfnamefont {C.~R.~I.}\ \bibnamefont {Carceller}}, \bibinfo
  {author} {\bibfnamefont {J.~B.}\ \bibnamefont {Brask}},\ and\ \bibinfo
  {author} {\bibfnamefont {J.}~\bibnamefont {Bae}},\ }\href
  {https://doi.org/10.1103/PRXQuantum.3.030337} {\bibfield  {journal} {\bibinfo
   {journal} {PRX Quantum}\ }\textbf {\bibinfo {volume} {3}},\ \bibinfo {pages}
  {030337} (\bibinfo {year} {2022})}\BibitemShut {NoStop}%
\bibitem [{\citenamefont {Schmid}(2024)}]{Schmid2024reviewreformulation}%
  \BibitemOpen
  \bibfield  {author} {\bibinfo {author} {\bibfnamefont {D.}~\bibnamefont
  {Schmid}},\ }\href {https://doi.org/10.22331/q-2024-01-03-1217} {\bibfield
  {journal} {\bibinfo  {journal} {{Quantum}}\ }\textbf {\bibinfo {volume}
  {8}},\ \bibinfo {pages} {1217} (\bibinfo {year} {2024})}\BibitemShut
  {NoStop}%
\bibitem [{\citenamefont {Gross}(2006)}]{gross2006hudson}%
  \BibitemOpen
  \bibfield  {author} {\bibinfo {author} {\bibfnamefont {D.}~\bibnamefont
  {Gross}},\ }\href {https://aip.scitation.org/doi/10.1063/1.2393152}
  {\bibfield  {journal} {\bibinfo  {journal} {J. Math. Phys.}\ }\textbf
  {\bibinfo {volume} {47}},\ \bibinfo {pages} {122107} (\bibinfo {year}
  {2006})}\BibitemShut {NoStop}%
\bibitem [{\citenamefont {Cavalcanti}\ and\ \citenamefont
  {Lal}(2014)}]{cavalcanti2014modifications}%
  \BibitemOpen
  \bibfield  {author} {\bibinfo {author} {\bibfnamefont {E.~G.}\ \bibnamefont
  {Cavalcanti}}\ and\ \bibinfo {author} {\bibfnamefont {R.}~\bibnamefont
  {Lal}},\ }\href {https://doi.org/10.1088/1751-8113/47/42/424018} {\bibfield
  {journal} {\bibinfo  {journal} {Journal of Physics A: Mathematical and
  Theoretical}\ }\textbf {\bibinfo {volume} {47}},\ \bibinfo {pages} {424018}
  (\bibinfo {year} {2014})}\BibitemShut {NoStop}%
\end{thebibliography}%


\end{document}